\documentclass[aps,prb,twocolumn,amsmath,showpacs]{revtex4} 

\usepackage{dcolumn}
\usepackage{bm}
\usepackage{color,graphicx}
\usepackage{amsfonts}
\usepackage{amssymb}
\usepackage{rotate}

\usepackage{subfigure}
\usepackage[latin1]{inputenc}
\usepackage{subfigure}
\usepackage{amsmath}
\usepackage{dcolumn}

\newcommand{\derivX}[1]{\ensuremath{\frac{\partial}{\partial #1}}}
\newcommand{\derivXN}[2]{\ensuremath{\frac{\partial^{ #2}}{\partial #1^{#2}}}}

\newcommand{\br}{{\mathbf{r}}}
\newcommand{\bk}{{\mathbf{k}}}

\newcommand{\be}{\begin{equation}}
\newcommand{\ee}{\end{equation}}

\renewcommand{\-}{\,-\,}

\def\oh{\frac{1}{2}}

\def\rf#1{(\ref{#1})}

\let\oldmarginpar\marginpar
\renewcommand\marginpar[1]{\-\oldmarginpar[\raggedleft\tiny #1]%
{\raggedright\tiny #1}}

\begin{document}

\title{Structure and consequences of vortex-core states in $p$-wave superfluids}

\author{G.~M\"{o}ller,$^{1,2}$ N.~R.~Cooper$^1$ and V.~Gurarie$^2$}
\affiliation{$^1$TCM Group, Cavendish Laboratory, J.~J.~Thomson Ave., Cambridge CB3 0HE, UK\\
$^2$University of Colorado at Boulder, Duane Physics UCB 390, Boulder CO 80309}
\date{January 21, 2011}


\begin{abstract}
We study the properties of the subgap states in $p$-wave superfluids, which occur at energies below the bulk gap and are localized inside 
the cores of vortices. We argue that their presence affects the topological protection of the zero modes. Transitions between the subgap states, 
including the zero modes and at energies much smaller than the gap, can alter the quantum states of the zero-modes. 
Consequently, qubits defined uniquely in terms of the zero-modes do not remain coherent, while compound qubits involving the zero-modes
and the parity of the occupation number of the subgap states on each vortex are still well defined. 
In neutral superfluids, it may be difficult to measure the parity of the subgap states. We propose to avoid this difficulty by working in the
regime of small chemical potential $\mu$, near the transition to a strongly paired phase, where the number of subgap states is reduced. 
We develop the theory to describe this regime of strong pairing interactions and we show how the subgap states are ultimately absorbed into the bulk gap.
Since the bulk gap also vanishes as $\mu\to 0$ there is an optimum value $\mu_c$ which maximises the combined gap.
We propose cold atomic gases as candidate systems where the regime of strong interactions can be explored, 
and explicitly evaluate $\mu_c$ in a Feshbach resonant  $^{40}$K gas. In particular, the parameter $c_2$ parametrizing 
the strength of the resonance in such gases, sets the characteristic size of vortices, and the energy scale of the subgap
states.
\end{abstract}

\maketitle

\section{Introduction}

Since the vision of a quantum computer based on the enigmatic degrees of freedom of topological phases 
of matter was set out,\cite{Kitaev2003} the search for physical realizations of topological phases has evolved as a leading
topic of condensed matter physics.\cite{TQCReview} The physics of chiral $p_x+ip_y$ BCS pairing\cite{ReadGreen} represents a simple prototype system
for topological order with prospective applications for inherently fault-tolerant topological quantum computing.\cite{TQCReview}

The $p_x+ip_y$ paired phase is believed to occur in a number of settings, including in the $A1$-phase of superfluid $^3$He,\cite{VolovikHe,KopninSalomaa}
in the bulk of the layered perovskite oxide Sr$_2$RuO$_4$,\cite{RiceSigrist,IshidaSrRuO}
as well as in two-dimensional samples of cold atomic\cite{GurarieRadzihovskyPRL,Botelho05,ChengYip,KlinkhammerVolovik,GurarieRadzihovsky} and polar molecular\cite{CooperShlyapnikov}
gases, and as a very closely related $p$-wave superconductor of composite fermions in the $\nu=5/2$ quantum Hall 
effect.\cite{Willett87, Xia2004,MooreRead91,NayakWilczek96,ReadGreen} 
Physics similar to that of the topologically non-trivial Majorana modes in $p_x+ip_y$ superconductors may also 
be induced by the proximity effect in interfaces between $s$-wave superconductors and topological insulators,\cite{FuKaneProximity2008} 
or related semiconductor heterostructures.\cite{SauProximitySC2010,AliceaSandwich}

In $p$-wave superfluids, topologically non-trivial degrees of freedom arise from the Majorana zero energy modes (ZEM) localized in
vortices of the superfluid order parameter.\cite{Ivanov} The topological protection required for the ZEM to be used in quantum computing relies 
on the existence of an energy gap towards quasiparticle excitations. It is troubling therefore, that vortex cores in superfluids feature eigenstates 
occurring at energies much smaller than the bulk gap.\cite{Caroli1964,KopninSalomaa}

In this paper, we discuss the implications of the presence of such subgap states for implementations of topological quantum 
computing (TQC) in a $p_x+ip_y$ paired phase.
We conclude that, while these states do not necessarily lead to decoherence of quantum
information, they can complicate significantly the construction of any practical scheme for TQC. 
In brief, at temperatures above the energy of the lowest subgap state $\epsilon_1$, thermal excitations of the systems include
processes which correspond essentially to a random flip of the qubit associated with the ZEM, represented by matrix-elements
involving a \emph{single} Majorana operator. However, as long as no excitations above the bulk gap are created, no decoherence
may occur, and information remains local to the vortex. Therefore, a compound qubit consisting of the ZEM as well as the complete 
set of subgap states is still well defined, with its state determined by the ZEM as well as the parity of the number of subgap excitations.

The requirement to perform measurements of such compound qubits, however, may prevent implementations of TQC in practice, 
particularly in neutral superfluids where interferometry is not applicable. As
a possible solution to this dilemma, we suggest the use of spin-polarized atomic Fermi gases which may be driven to a strongly interacting
regime of $p_x+ip_y$ pairing in a controlled fashion (remaining in the weak-pairing phase\cite{ReadGreen}), by exploiting the physics of the BEC-BCS 
crossover\cite{NozieresSchmidtRink} in cold atomic gases.\cite{Greiner2003,GurarieRadzihovsky} 
This regime, where the bulk gap can be of the
order of the Fermi energy was not considered in detail in previous studies of vortex states in the $p_x+ip_y$ paired 
phases.\cite{KopninSalomaa,Mizushima2008,KrausPRL,KrausPRB} In particular, the existing scheme leading to an approximate analytical solution
for the subgap states\cite{KopninSalomaa} does not apply.

Following the need to elucidate the strongly paired regime of neutral superfluids, in this paper we establish the theoretical framework for
characterizing the physics of a quasi two-dimensional cold atomic gas in the BCS regime above a $p$-wave Feshbach resonance. 
We show how to describe the system in a two-channel
model, and deduce the connection to the underlying parameters of the atomic gas, studying the example of $^{40}$K, in particular.
As a core result of this analysis, the size of the vortex cores is set by the parameter $c_2$ that characterizes the strength of the resonance.
We study the ensuing Bogoliubov de Gennes (BdG) equations in the limit of strong interactions and show how the subgap states merge with the
bulk gap one by one until none is left. We then focus particularly on the nature of the first excited subgap state, as opposed to a number of
previous studies that focused on the zero modes.\cite{MengSplitting09,KrausPRL,SauStability,SilaevVolovik,MizushimaMachida10a,MizushimaMachida10b}
Based on a numerical study of the BdG equations, we identify the regime with the best prospects for realizing TQC in $p$-wave 
superfluid cold atomic gases, i.e., the regime maximising the energy of the first subgap state $\epsilon_1$ as a function of the detuning.

The structure of this paper is as follows: in section \ref{sec:model}, we review the Bogoliubov de Gennes equations for the
description of $p$-wave superfluids, identify the chemical potential as the single parameter of these equations in suitable 
rescaled units, and restate some known results about the subgap states in these units. Section \ref{sec:role} provides an in-depth
discussion of how the subgap states influence the topological protection of the manifold of ZEM in a system with many vortices.
We then proceed in section \ref{sec:strongInteractions} to discuss the theoretical framework for the description of atomic $p$-wave superfluids in 
the regime of strong interactions, using a two-channel model. Solutions to the ensuing Bogoliubov de Gennes equations are given in Section \ref{sec:solutions},
mostly based on a numerical study with a strong focus on the properties of the first excited subgap state. Finally, in section \ref{sec:exp_consequences}
we deduce concrete numbers for experimental realizations of a $p$-wave superfluids based on a Feshbach resonance in potassium gases, before
presenting further conclusions in section \ref{sec:conclusions}. Details of how to extract the physical parameters for the Feshbach resonance in $^{40}$K are
given in Appendix \ref{sec:experimentalC2}.
 Appendix \ref{sec:subgapAnalytic} is devoted to a review of the approximate analytic calculation of the subgap states. 
 Appendix \ref{sec:matrix_elements} provides a discussion of the matrix elements between subgap states, and in
Appendix \ref{sec:BdGSphere} we review how to express the BdG equations on the sphere.

\section{Model}
\label{sec:model}
Before entering the main discussion of this paper, let us introduce the notations of the formalism that we make use of below.
The spectrum of spinless (fully spin-polarized) fermions whose $p_x+ip_y$ pairing order parameter is described by a gap function
$\Delta(\br)$ and gives rise to the Bogoliubov equations
\begin{equation}
\label{eq:BdG}
  \left(
    \begin{array}{cc}
      \hat h-\mu & \hat \pi \\
      \hat \pi^\dagger &
      -\hat h^T + \mu 
    \end{array}
  \right)  
  \left(
    \begin{array}{c}
      u_{n}\\
      v_{n}
    \end{array}
  \right)
  =
  E_{n}
  \left(
    \begin{array}{c}
      u_{n}\\
      v_{n}
    \end{array}
  \right),
\end{equation}
where, in a coordinate representation, the single particle kinetic term can be expressed as $\hat h=-\frac{\nabla^2}{2m}$ 
and the pairing term as $\hat \pi=\frac{1}{2}\sqrt{\Delta(\br)} \left(\derivX{x}-i\derivX{y}\right)\sqrt{\Delta(\br)}$.
Note that in this definition of $p$-wave pairing, the gap function $\Delta$ has units of Energy$\times$Length 
in contrast to $s$-wave superconductors. Below, we adopt a dimensionless version of 
the Bogoliubov equations, obtained from (\ref{eq:BdG}) by rescaling the equations in terms of the characteristic
length set by the gap function
\begin{subequations}
\label{eq:rescalings}
\begin{equation}
   L=\frac{1}{m\Delta_0},
\end{equation}
yielding the dimensionless length and energy-scales
\begin{equation}
  \bar x = \frac{x}{L}\text{, and } \bar E = EmL^2=\frac{E}{m\Delta_0^2}.
\end{equation}
\end{subequations}
Note the scale $L$ is determined by the asymptotic value
of the gap function in the bulk, $\Delta_0$, measured far away from any vortices. 
We should note that while the gap function $\Delta_0$ yields a characteristic energy scale 
$E_0=m\Delta_0^2$ of the problem, this scale is distinct from the bulk gap, as defined below, 
that represents a second useful reference energy.
For the remainder of this paper, we use this dimensionless formulation of the BdG equation (\ref{eq:BdG}),
which is formally equivalent to setting the mass $m=1$ and the value of the gap function in the bulk $\Delta_0=1$.
This leaves $\bar \mu = \mu/[m\Delta_0^2]$ as the single dimensionless parameter of the problem. In the remainder
of this paper, symbols with a bar refer to the dimensionless versions, while bare symbols represent fully dimensional quantities.
Wherever dimensionless parameters appear in an equation, the other parameters are also dimensionless even if not
explicitly indicated as such.

Let us indicate a few known results in the dimensionless units. For example, the dispersion of the bulk quasiparticle 
excitations in the absence of any vortices now reads
\begin{equation}
\bar E_\bk = \sqrt{(k^2/2-\bar\mu)^2+k^2}.
\end{equation}
The bulk gap $\bar\Delta_B$ is set by the minimum of $E_\bk$ which occurs at momentum at $k=0$ or at $|k|=\sqrt{2(\bar\mu-1)}$ 
for $\bar\mu< 1$ or $\bar\mu > 1$ respectively, with
\begin{equation}
\label{eq:bulk_gap}
\bar\Delta_B=\left\{
\begin{array}{ll}
  \bar\mu,&\text{for }\bar\mu\leq 1 \\
\sqrt{2\bar\mu-1}, &\text{for }\bar\mu>1 \\
\end{array}
\right.
\end{equation}
Our study focuses on the spectrum of the $p_x+i p_y$ superconductor in the presence of a radially
symmetric vortex described by a winding of the order parameter according to $\Delta(\br) = h^2(r) e^{i\kappa \phi}$,
where $h(r)\to 1$ at large $r$.
Setting $u(\br)=\exp [i(m+\frac{\kappa-1}{2})\phi]u(r)$ and $v(\br)=\exp [i(m-\frac{\kappa-1}{2})\phi]v(r)$, the Bogoliubov equations 
for general $\kappa$ read:
\begin{align}
\label{eq:BdGVortex}
-\left[\mathcal{D}_{m+\frac{\kappa-1}{2}}+\bar\mu\right] u + h^2 \left( \derivX{r} + \frac{\frac{1}{2}+m}{r} \right)v + h h' v = \bar E u \nonumber\\
 \left[\mathcal{D}_{m-\frac{\kappa-1}{2}}+\bar\mu\right] v - h^2 \left( \derivX{r} + \frac{\frac{1}{2}-m}{r} \right)u - h h' u = \bar E v
 \end{align}
where we introduced an abbreviation for the second order differential operator 
\begin{equation}
\mathcal{D}_l = \frac{1}{2} \derivXN{r}{2} + \frac{1}{2r} \derivX{r} - \frac{l^2}{2r^2}.
\end{equation}
In this paper, we focus on the cases of a vortex ($\kappa=1$) and antivortex ($\kappa=-1$). 
Analytical solutions to the equations (\ref{eq:BdGVortex}) are known for several regimes of parameters.\cite{Caroli1964,KopninSalomaa} 
Most importantly, a zero mode exists for all vortices of odd vorticity, and is topologically protected against perturbations
that do not destroy the bulk gap.\cite{Tewari07, 
GurarieRadzihovskyPRB, KrausPRB}
In a cylindrically symmetric vortex of vanishing size\cite{GurarieRadzihovskyPRB} and with vorticity $\kappa=2n-1$, the wavefunction of the
zero-mode takes the form
\begin{equation}
  \label{eq:zero_mode}
  u(r) = v(r) = \left\{
    \begin{array}{ll}
       J_n\left( r \sqrt{2\bar\mu-1} \right) e^{-r}, &\text{for }\bar\mu\geq\frac{1}{2} \\
        I_n\left( r \sqrt{1-2\bar\mu} \right) e^{-r}, &\text{for }\bar\mu<\frac{1}{2}
    \end{array}
  \right.,
\end{equation}
with the (modified) Bessel functions $J_n$ ($I_n$) [and the angular dependency involves and appropriate phase factor as introduced above]. 
If a finite sized vortex core is considered, both the Bessel function and the exponential localization are modified. The latter instead becomes $\exp[ -\int_0^r h^2(r')dr']$.

It is also known that there may exist additional subgap states,\cite{Caroli1964,KopninSalomaa} 
with finite $E< \Delta_B$  known as the Caroli-deGennes-Matricon (CdGM) states.\cite{Caroli1964}  
While the energy of the zero-mode is protected, the energy of the CdGM states depends on the shape of the vortex. 
Their energy was calculated in the limit of $\bar\mu\gg 1$ for the closely related case of a $\kappa=1$ vortex in the 
A phase of $^3$He,\cite{Caroli1964,KopninSalomaa} and found to be
\begin{equation}
  \label{eq:subgap_energy}
  \bar E_m=-m\bar \omega_0
\end{equation}
for the subgap state with angular momentum $m$, as long as $E\ll \Delta_B$ and with
\begin{equation}
  \label{eq:subgap_spacing}
  \bar \omega_0 = \frac{\int_0^\infty dr \frac{h^2(r)}{r}\exp\left[-2 \int_0^r dr' h^2(r')\right]}
  {\int_0^\infty dr \exp\left[-2 \int_0^r dr' h^2(r')\right]}.
\end{equation}
Note the spacing of the subgap states is independent of the dimensionless $\bar \mu$, whereas in dimensional units 
$\omega_0\sim\Delta_B^2/\mu\ll\Delta_B$. \footnote{This full expression can be recovered by dimensional analysis:
Units of energy are $E=[m\Delta_0^2]$, and $\Delta_B\sim\sqrt{\bar\mu}$. Thus, $\Delta_B=m\Delta_0^2\sqrt{\mu/(m\Delta_0^2)}=\sqrt{m}\Delta_0\sqrt{\mu}$
and $\omega_0\sim \Delta_B^2/\mu$.}
There are thus $\mu / \Delta_B \sim \sqrt{\bar\mu}$ such modes
for large $\bar\mu$. Below, we consider the behaviour in the limit of small $\bar\mu$ and find that as $\bar\mu$ is reduced, 
the subgap modes merge with the bulk one by one. 
For completeness, the perturbative solution for these eigenstates is included as Appendix \ref{sec:subgapAnalytic}. 
Finally, we introduce $\epsilon_1$ as a notation for the energy of the first subgap state, with
$\epsilon_1\to\omega_0$ in the limits of the approximation of large $\bar\mu$.

\section{Role of the subgap states}
\label{sec:role}
Given the presence of states below the gap energy, the question arises how these states affect the topological protection
of the Majorana zero-modes. The conventional view of the topological protection of a groundstate manifold requires the
existence of an energy gap towards excited states.\cite{Kitaev2003} At sufficiently low temperatures, the probability of
creating an excitation is then exponentially suppressed. If the vortices holding zero-modes are well separated from each other, recombination 
would most likely occur in a way that restores the original groundstate configuration: due to the topological nature of the
system, a quasiparticle excitation can only permanently change the state of the system if it braids around a second 
vortex.\cite{TQCReview} Excitations of the subgap states always remain localized to a single vortex. However, even if they cannot 
propagate information to a second vortex, they can make reading off the state of the qubit very complicated, as we see later. 

For $p$-wave superconductors in the BCS limit, the first subgap energy $\epsilon_1$ is typically much smaller than the 
bulk gap $\Delta_B$ (by a factor of $\Delta_B/\mu$), and it may thus be impracticable or even impossible to cool the system 
to temperatures below $\epsilon_1$. For temperatures above $\epsilon_1$, excitations of the subgap state(s) are likely.

Transitions between the zero-modes and the subgap states require non-zero matrix elements connecting these states.
It can be shown that matrix-elements for a scalar potential, created for instance by a passing phonon, are indeed non-zero
by expanding a disorder potential $\hat V$ in the basis of Bogoliubov eigenstates.
Therefore, transitions into the subgap states will occur at a finite rate in thermal equilibrium. In this case, the necessary energy for 
quantum jumps is then supplied by the heat bath, for example in the form of phonons in the ruthenates. 
An explicit calculation of the matrix elements involving the zero modes is shown in Appendix \ref{sec:matrix_elements}.
For the purpose of this discussion, we only need to acknowledge the presence of processes of the form 
$\hat c_\nu^{\dagger}\hat \gamma_\nu$ or $\hat \gamma_\nu\hat c_\nu$, involving Majorana fermions $\hat \gamma_\nu$
and fermionic subgap states $\hat c_\nu^{(\dagger)}$ at vortex $\nu$.

For any static external potential, the system has well-defined eigenstates with infinite lifetime. 
Time-dependent scalar potentials however, can provide the energy to make transitions between eigenstates.
The zero-mode has an interesting property in that its \emph{energy} is protected against perturbations.
However, its \emph{wavefunction} is deformed by a changing potential. Thus, the zero mode for a given external 
potential has a non-zero overlap with excited states of an evolving potential at a later time. Non-adiabatic transitions can only be induced by  
perturbations, which occur sufficiently quickly on a time-scale set by the energy-scale for transitions, as the
required energy is supplied by the force resulting from a time-dependent potential. Therefore, the presence of 
the subgap states sets tighter limits on how stationary the scalar potential needs to be to conserve adiabaticity.
In this context, we note that even a stationary disorder potential will act as a time-dependent perturbation driving 
non-adiabatic processes under braiding of the vortices.

Having identified possible non-adiabatic processes among the zero modes and subgap states, the question 
remains whether these transitions cause decoherence in the system.
Let us first assume that we are able to momentarily cool the system to very low temperatures, even though 
non-adiabaticity cannot be avoided during braiding operations. Thus, we may start and end in the groundstate.
The consequences of temporary transitions to the subgap states can be analyzed
using the operator describing the braid which exchanges the positions of two vortices $\hat\gamma_1$ 
and $\hat\gamma_2$, and which is given by $\beta_{12}=\frac{1}{\sqrt{2}}(1+\hat\gamma_1\hat\gamma_2)$ 
(followed by a renaming of the vortices).\cite{Ivanov}

For instance, excitation of the subgap state on vortex 1, followed by an interchange of 1 and 2 and de-excitation
of the vortex is equivalent to a simple exchange up a to sign:
\begin{align}
\hat c_2 \hat \gamma_2 (1+\hat \gamma_1\hat\gamma_2)\hat\gamma_1 \hat c_1^\dagger = - ( 1+ \hat\gamma_1\hat\gamma_2)
\end{align}
Generally, the non-adiabatic processes affect braiding processes only up to a sign (or rather, up to a random abelian phase 
caused by the lack of knowledge of the energy as a function of time). More generally, any even number of transitions
to subgap states will lead to an even number of Majorana operators being inserted -- thus leaving the final state invariant.
This is always the case if both the initial and the final state are within the groundstate manifold.
However, should an odd number of subgap states be excited on a vortex, an error occurs and the final quantum state is altered [changing
the entanglement properties of the wavefunction by modifying the relative phases of the components with empty and occupied core-states\cite{SternOppen}].

Provided that all subgap states remain local to its vortex, and provided one can measure the number of fermions in the
subgap states of all vortices, one can deduce how many Majorana operators must have been inserted in the initial state, and correct
the final state accordingly. While such a procedure seems fundamentally possible, it would likely be extremely inconvenient in practice.

It may be possible to find a measurement scheme which evaluates the state of the Majorana mode as well as the parity of
the subgap states, which could be taken altogether as a compound qubit. Such qubits were considered independently in a recent 
paper.\cite{AkhmerovParity} We now discuss the feasibility of measuring the atom number parity in
the context of possible read-out schemes for quantum bits in $p$-wave superconductors / superfluids. 

As an example, let us consider a proposal to detect the state of qubits in atomic $p$-wave superfluids via spectroscopy.\cite{TewariDasSarma07}
This scheme relies on the possibility to detect the presence or absence of a single unpaired atom after fusing two vortices forming a qubit.
Detecting a single unpaired atom is possible as the coupling of the atom's internal states to a Raman pulse depends sensitively on the
detuning of the incident radiation. However, the required detuning differs by twice the bulk gap between paired and unpaired atoms, 
enabling one to address only the latter. Given the presence of the subgap states, it is now possible that there are multiple unpaired atoms
in a single vortex. Consider, for example, a situation where an even number of subgap states have been excited in the two vortices of the qubit.
Then, according to the proposal under consideration,\cite{TewariDasSarma07} we would detect the presence of unpaired atoms; however,
the state of the qubit corresponds to an empty $|0\rangle$ state. To deduce the state of the qubit, it is therefore required to detect the
\emph{number} of unpaired fermions (which equals the sum of occupation numbers of all subgap states) to be able to deduce its the qubit state 
from the parity. This requires more accurate detection techniques.

Interferometry appears as a suitable measurement method, as it is sensitive only to the parity of the number of atoms. In
\emph{charged} superfluids, interferometry is expected to work,\cite{Kitaev2003} provided that transport along the edge
can be described in terms of the motion of quasiparticles with no internal degrees of freedom.
However, in compressible \emph{neutral} $p$-wave superfluids, interferometry is certainly not applicable at all as the compressibility
of the system entails particle number fluctuations that wash out the interferometric signal when braiding quasiparticles.\cite{HaldaneWu85} 
For incompressible states such as the Moore-Read phase found in the $\nu=5/2$ quantum Hall effect, interferometry is considered 
as a leading scheme for read-out;\cite{TQCReview} we should also note in passing that there are no subgap states in the spectrum for 
FQHE states, as the size of the vortex is of the order of the interparticle spacing.

As the presence of the subgap states gives additional structure to the quantum bits, we anticipate that the read-out protocols will be more 
complex, and thus be prone to errors in the read-out process, particularly for neutral superfluids. Therefore, the presence of low-lying subgap
states very likely prevents the practical use of the quantum register in these systems.

\section{$P$-Wave superfluids in the limit of strong interactions}
\label{sec:strongInteractions}
In the previous section, we argued that the presence of low lying subgap states complicates the use of $p$-wave
superfluids for TQC. From a technological point of view it seems imperative to evade the problem of having to keep 
track of the occupation numbers of all subgap states. In $p_x+ip_y$ superfluids of cold atomic gases, this can
be achieved by driving the system into the regime of strong interactions where the number of subgap states $N_s$ is expected 
to be small. For $\bar\mu\gg 1$, the semiclassical approximation\cite{KopninSalomaa} yields $N_s\propto \sqrt{\bar\mu}$.
The core task of this paper consists in establishing the nature of the subgap states for $\bar\mu$ small.

To do so, we need to solve the BdG equations for a vortex when $\bar\mu$ is small. This can only be done numerically. The appropriate 
equations depend on the shape of the condensate $h(r)$, which vanishes linearly in $r$ near the core for singly quantized vortices. 
At the same time, as $r$ increases $h(r)$ approaches $1$ on a characteristic lengthscale called the coherence 
length of the condensate, $\xi$. We need to know $\xi$ to fully define the BdG equations.

A known way to create $p$-wave superconductors with $\bar\mu$ which is not very large is by employing cold atomic gases with 
$p$-wave Feshbach resonances. Current experimental realizations of atomic gases with $p$-wave Feshbach resonances are not 
stable,\cite{RegalPWave03, FuchsPWave08, InadaPWave08} due to 3-body recombination.\cite{Castin08, LevinsenPRA}
However, there are proposals to enhance stability using optical lattices.\cite{SyassenStability,HanZoller09}
In the next subsection we study such systems in order to determine its parameters, including the 
scale of the gap function, the chemical potential and the coherence length of the condensate.
Additionally, we also consider $p$-wave superfluids of dipolar molecules in section \ref{sec:dipolar}.

\subsection{Atomic Fermi gases}

In the context of cold atomic gases, $p$-wave pairing is studied via the two channel model that considers
free fermions and their bosonic bound states as separate entities.\cite{GurarieRadzihovsky}
Its Hamiltonian reads
\begin{eqnarray} \label{eq:hamil}
 H &=& \sum_{\bf p} \frac{p^2}{2m} ~\hat a^\dagger_{\bf p}
\hat a_{\bf p} + \sum_{{\bf q}, \alpha} \left(\epsilon_0 +
{\frac{q^2}{4m}} \right)
\hat b_{\alpha{\bf q}}^\dagger \hat b_{\alpha {\bf q}}\cr
&+&\sum_{{\bf p},{\bf q},\alpha} ~\frac g  {\sqrt{V}} \left( \hat b_{\alpha
{\bf q}} ~p_\alpha~ \hat a^\dagger_{\frac{\bf q}  2+{\bf p}} ~\hat a^\dagger_{ \frac {\bf q}
 2-{ \bf p}} + h. c. \right).
\label{eq:H}
\end{eqnarray}
This model describes a gas of fermions with creation and annihilation operators $\hat a^\dagger$, $\hat a$ which can form bosonic molecules with the angular 
momentum $1$ (hence the bosonic molecules are described by the creation and annihilation operators $\hat b^\dagger_\alpha$, $\hat b_\alpha$ with the vector 
index $\nu$). Once the bosonic molecules Bose-condense, the fermions form a $p$-wave superconductor. For homogeneous $p$-wave superfluids, it was found 
that the $p_x+i p_y$ paired phase is always the groundstate.\cite{GurarieRadzihovskyPRL,ChengYip} In the experimentally relevant case of gas with a dipolar
anisotropy, the triplet of $m=+1,0,-1$ states is split, and the $m=0$ state becomes the groundstate, and as a result $p_x$-pairing is present in the phase diagram. 
For weak anisotropies, the chiral state remains the low-temperate phase over a large range of detunings. Even in the case of strong splitting, the chiral $p_x+i p_y$
paired phase remains present: Feshbach resonances for the $m=0$ and $m=\pm 1$ channels are then also well separated, and the chiral phase is observed
near the latter one.\cite{GurarieRadzihovsky}

We observe that in real experiments, the cold atomic gas would be confined to two dimensions up to a ``pancake" of width $\ell$ of the order of the wavelength of visible light, or $500$nm. The momenta in Eq.~(\ref{eq:H}) are chosen appropriately to reflect such a geometry. We call this setup a quasi-2D gas (unlike a purely 2D setup where all motion is completely confined to 2D geometry). 

The confinement length $\ell$, while much smaller than interparticle separation, thus leading to a truly 2D superconductor,  is still much larger than the molecular size $R_e$, which is set by the range of the forces responsible for the formation of the molecules, typically about $1-3$nm. Therefore the formation of molecules remains a 3D process. 

The detuning $\delta$, the parameter which can be controlled experimentally by varying the magnetic field, allows to vary the strength of pairing between the fermions. $\delta$ is simply related to the ``bare detuning" $\epsilon_0$, the parameter which appears explicitly in the Hamiltonian, by the relation
\begin{equation} \label{eq:rendet} \delta = \frac{\epsilon_0-{\rm const}}{1+c_2}, \end{equation}
where const is an irrelevant constant, and $c_2$ is a parameter which characterizes the strength of the resonance, 
\begin{equation} c_2 = \frac{m^2}{3\pi^2} g^2 \Lambda  = \frac{m^2}{3\pi^2} \frac{g^2}{R_e}.
\end{equation}
The parameter $\Lambda=1/R_e$ is hidden in the two-channel model, appearing as an upper cutoff for all the sums over momenta. 

The solution to the two-channel model can be described in the following way.  First of all, the bosons are always Bose condensed, or 
\begin{equation} \label{eq:bosons} \left< {\hat b_{\alpha, {\bf q=0}}} \right> = \sqrt{V} B_{\alpha}.
\end{equation}
This leads to fermions forming a $p$-wave superconductor. 
Second, the vector structure of $B_\alpha$ defines the order parameter, and for a $p_x+ip_y$ pairing, energetically favorable within this model, 
\begin{equation}
B_x=-iB_y=B.
\end{equation} 
Given the density of particles $n$ and the corresponding Fermi energy $\epsilon_F$ (understood as the Fermi energy of a free Fermi gas at this density), we can identify three regimes. 

If $\delta>2 \epsilon_F$, then the density of bosons is exponentially small\cite{GurarieRadzihovsky} in the parameter $\exp\left\{- \frac{\delta-2\epsilon_F}{\epsilon_F}\mathcal{S}\right\}$, where $\mathcal{S}$ is defined below in Eq.~(\ref{eq:scaleS}).
Since the density of bosons 
\begin{equation} n_b = 2B^2
\end{equation}
is responsible for the superconducting pairing in this problem via
\begin{equation} \Delta_0 = 2 g B \end{equation}
the regime of exponentially small $n_b$ corresponds to the conventional BCS superconductors in which the transition temperature 
is a small fraction of the Fermi energy. The chemical potential in this regime is $\mu = \epsilon_F$.

Next, if $0<\delta<2 \epsilon_F$, a finite fraction of fermions converts into bosons. $n_b$ is now of the order of the initial density of fermions in this problem, and the superconductor which forms under these conditions has a transition temperature which is a substantial fraction of the Fermi energy (and growing as $\delta$ is decreased). At the same time, the chemical potential of the fermions $\mu$ is approximately equal to $\delta/2$, 
\begin{equation}
\mu \approx \frac{\delta}{2},
\end{equation}
where ``approximately" means up to terms of the order of $g^2$). 
Thus in the terminology of Ref.~\onlinecite{ReadGreen} this is still a weakly paired superconductor (that is, a superconductor with $\mu>0$ which has large Cooper pairs).  
We refer to this regime as strongly interacting.

Finally, when $\delta$ becomes negative, the chemical potential changes sign. Now the density of fermions is exponentially small, while most particles are bosonic molecules which are Bose condensed. The transition temperature of such a superconductor is a certain finite fraction of the Fermi energy. In the terminology of Ref.~\onlinecite{ReadGreen} this is now a strongly paired superconductor, separated from a weakly paired superconductor by a quantum phase transition which occurs at $\mu=0$ or $\delta$ close to 0.

Returning to the discussion of possible experimental realizations of two dimensional gases with cold atoms, the two-channel model describes a two dimensional superconductor which, at $\mu>0$ (implying positive $\delta$), is the topological state of matter of interest to us, here. A typical experiment would be conducted in the regime where $0<\delta<2\epsilon_F$ to maximize the transition temperature of the superconductor, and as we show later, to remove the subgap states for a particular $\delta$. In this regime, $0 <\mu <\epsilon_F$, unlike the conventional superconductors where $\mu=\epsilon_F$. The principal	 goal of this paper is to understand what happens to the vortex subgap states as the chemical potential becomes smaller than $\epsilon_F$.

We now use this scenario of the quasi-2D two-channel model to estimate the coherence length of the condensate $\xi$, which we need to be able to estimate the typical size of a vortex. 

To do that, we integrate out the fermions and concentrate on the effective action of bosons. This was done in Ref.~\onlinecite{GurarieRadzihovsky} with the result, in the regime where $\mu<\epsilon_F$,
\begin{eqnarray} \label{eq:effact} S &=& \int dV dt \left[ \bar b_\alpha \left( i \frac \partial {\partial t}-\frac{\Delta}{4m} + \delta -\mu \right) b_\alpha \, (1+c_2) +  \right. \cr && \left.  C g^2 c_2 m \left(\left( \bar b_\alpha b_\alpha \right)^2 +\frac{1}{2} |b_\alpha^2|^2 \right) \right].
\end{eqnarray}
Here $C$ is a dimensionless constant whose precise value is not important for our purposes here. We note that the calculations in Ref.~\onlinecite{GurarieRadzihovsky} are done in 3D, while we work in quasi-2D. However, the contributions from integrating out the fermions in (\ref{eq:effact}) are all proportional to $c_2 \sim g^2 \Lambda$. In other words, they come from the momenta of the order of $\Lambda \sim 1/R_e \gg 1/\ell$ where $\ell$, the width of the condensate, is much larger than $R_e$, the range of the interactions. So Eq.~(\ref{eq:effact}) is valid in quasi-2D, as well as in 3D. 

The coherence length $\xi$ can be extracted by comparing the kinetic and quartic terms of Eq.~(\ref{eq:effact}). We find
\begin{equation} 
\label{eq:balanceQQ}
\frac{1+c_2}{m \xi^2}  \sim g^2 c_2 m B^2,
\end{equation} where we replaced $\sum_\alpha \bar b_\alpha b_\alpha$ by $B^2$ in the spirit of Gross-Pitaevskii equation. This gives
\begin{equation} \xi \sim \frac {\sqrt{1+c_2}} {gB m \sqrt{ c_2}}.
\end{equation}

Throughout this paper, however, we are interested in the dimensionless $\bar\xi$, expressed in units provided by $\Delta_0$ [the units of length given by $1/(\Delta_0 m)$, see Eq.~(\ref{eq:rescalings})]. In turn, the dimensionless coherence length can be found as
\begin{equation} \label{eq:coherence}
\bar\xi \equiv \xi \Delta_0 m \sim \frac {\sqrt{1+c_2}} {gB m \sqrt{ c_2}} m g B =  \sqrt{\frac{1+c_2}{c_2}}.
\end{equation}
Thus we find that if $c_2$ is large, the coherence length in our units is close to 1. If $c_2$ is small, the coherence length can be larger than 1. 

In the $p$-wave Feshbach resonance in $^{40}$K, $c_2$ can be shown to be around $14.4$ (see Appendix \ref{sec:experimentalC2}), and the 
coherence length is close to 1. However we do not know the value of $c_2$ for other $p$-wave Feshbach resonances, so we cannot make any 
assumptions about it beyond Eq.~(\ref{eq:coherence}). 

We can further use (\ref{eq:hamil}) to calculate how the bulk gap $\Delta_0$ and the chemical potential $\mu$ depend on the detuning $\delta$. 
To do this, we rely on the following arguments.

If $\delta>2 \epsilon_F$, then $\Delta_0$ is exponentially small. At the same time $\mu=\epsilon_F$. When expressed in units of $m \Delta_0^2/\hbar^2$, as we do throughout this paper, $\bar\mu \gg 1$. This is the regime of conventional superconductors. 

When $\delta$ is lowered below $2 \epsilon_F$, then $\Delta_0$ quickly starts to increase. Let us determine $\Delta_0$ as a function of $\delta$ in this regime.
As before, we work in the quasi-2D regime, where the $p$-wave gas is confined to a pancake of width $\ell$, such that $\ell$ is much smaller than the average particle separation. We write down the particle conservation condition
\begin{equation}
\oh \sum_p \left[ 1- \frac{\frac{p^2}{2m}-\mu}{\sqrt{\left(\frac {p^2}{2m} - \mu \right)^2 + \Delta_0^2 p^2}}  \right]+ 2 \cdot 2 B^2 V = N_{\rm total}.
\end{equation}
Here, $B$ is the condensate density originally defined in (\ref{eq:bosons}), and $\Delta_0 = 2 g B$. 
The summation over $p$ reflects the quasi-2D geometry of the $p$-wave gas and may not be straightforward to convert into 
an integral, as is usually done in these cases. However, we observe that the summation over $p$ is actually divergent at large
$p \sim \Lambda \gg 1/\ell$. This allows us to capture this divergence by introducing a 3D integral $\sum_p \rightarrow V \int d^3p/(2\pi)^3$. 
This was already done in Ref.~\onlinecite{GurarieRadzihovsky}, with the result
\begin{align} 
 & \oh \sum_p \left[ 1- \frac{\frac{p^2}{2m}-\mu}{\sqrt{\left(\frac {p^2}{2m} - \mu \right)^2 + \Delta_0^2 p^2}}  \right] 
  \cr + \;& 4 B^2 V \left(1 + c_2 \right) = N_{\rm total}.
\end{align}
In this equation, the summation over momenta $p$ is now restricted by $p \ll 1/\ell$, while the contribution of the leading 
divergence at larger momenta can be absorbed into the $B^2$ term.
The remaining summation over $p$ is purely two-dimensional. We can evaluate it by approximating the expression in the 
square brackets as a Fermi-Dirac step. This yields
\be \frac{V}{\ell} \frac{m \mu}{2\pi} + 4 B^2 V (1+c_2) = N_{\rm total}.
\ee
Note the volume $V$ is a three-dimensional quantity, which yields the corresponding 2D volume as $V/\ell$. 
To express our final results, let us write also
\begin{equation} x = \frac{\delta}{2 \epsilon_F} \approx \frac{\mu}{\epsilon_F}\end{equation}
which varies from $0$ to $1$ and measures detuning $\delta/2$ in the units of Fermi energy. In terms of this parameter,
\be B = \sqrt{\frac{\epsilon_F m (1-x)}{8 \pi \ell (1+c_2)}}.
\ee
Finally, this gives $\Delta_0=2g B$ as
\be \Delta_0 =  g \sqrt{\frac{\epsilon_F m (1-x)}{2 \pi \ell (1+c_2)}}.
\ee
For the purposes of this paper, we would like to compute $\mu=\delta/2=x \epsilon_F$ in the units of $m \Delta_0^2$. This gives
the dimensionless chemical potential
\be \label{eq:mudim} \bar\mu = \frac{\mu}{m\Delta_0^2} = 2 \pi (1+c_2) \frac{\ell}{m^2 g^2} \frac{x}{1-x}.
\ee
For future reference, we name the prefactor in this equation which sets the overall scale of the gap and chemical potential
\be \label{eq:scaleS} \mathcal{S} =2 \pi (1+c_2) \frac{\ell}{m^2 g^2}.
\ee
We can see that as the detuning is decreased past $2 \epsilon_F$, $x$ is varied from $1$ to $0$, the dimensionless $\bar\mu$ indeed varies from very large values [(\ref{eq:mudim}) predicts infinity at $x=1$, although this is an approximation artifact; at large detuning we are in the regime of conventional superconductor with a very large but finite  $\bar\mu$] all the way down to 0.

\subsection{Polar molecular Fermi gases}
\label{sec:dipolar}

For diatomic molecular gases in two dimensions, attractive interactions may be generated by dressing the molecules with circularly 
polarized microwave radiation.\cite{CooperShlyapnikov} The result is a dipole moment which is rotating in the plane of the 2D gas, and the
interaction averaged over the angle of rotation yields a net attractive long-range potential $V(r)=- d_\text{eff}^2/(2r^3)$. (We follow the notations
of Ref.~\onlinecite{CooperShlyapnikov}.)
The strength of this interaction depends on the field strength for the incident microwave interaction, as well as on the permanent dipole
moment of the molecules. It is characterized by a lengthscale $r^*=Md_\text{eff}^2/2\hbar^2$, that can be of the order of $r^*\simeq 200$nm
for realistic experimental parameters in typical $^7$Li$^{40}$K molecules. The dimensionless strength of the interaction $k_F r^*$, can therefore
be of order one in gases of a fairly low density.\cite{CooperShlyapnikov}

For superfluids with dipolar interactions, theory has been developed only at the level of a BCS mean-field description.\cite{CooperShlyapnikov}
In this framework, the critical temperature and bulk order parameter $\Delta_0$ are obtained to be
\be
\label{eq:Tc_dipolar}
T_c \sim \Delta_0 k_F \sim \epsilon_F e^{-\frac{3\pi}{4k_F r^*}}.
\ee
For large $k_Fr^*$, we expect that a significant fraction of the fermions will be paired, leading to a reduction of the chemical potential for fermions.
However, to quantify this effect more theory would have to be developed to study the specific case of dipolar interactions.

Similar to the reasoning leading to Eq.~(\ref{eq:balanceQQ}), the coherence-length $\xi$ is obtained by balancing the kinetic and interaction terms in
the underlying pairing Hamiltonian. The result can be expressed in terms of the Fermi-velocity and the critical temperature
\be
\label{eq:coherence_dipolar}
\xi \sim \frac{v_F}{T_c}\sim \lambda_F e^{+\frac{3\pi}{4k_F r^*}},
\ee
and when stated in our dimensionless units, this equates to a coherence length of order one
\be
\label{eq:coherence_dipolar_nodim}
\bar \xi = \xi m \Delta_0 = \lambda_F m \epsilon_F/k_F \simeq 1.
\ee
Note that generically, $\bar\xi\simeq 1$ for any superfluid in the BCS limit, as we have used no specific features of the dipolar interaction to derive this
relation.

\section{Solutions of the BdG equations}
\label{sec:solutions}
The two-channel model discussed in the previous section provides a framework to discuss the physics of molecule formation in strongly interacting superfluids.
Formally however, in strongly interacting $p$-wave superfluids with large $c_2$, one can integrate out the bosons in this problem and reduce the 
two-channel model to the one-channel model described by the usual Bogoliubov de-Gennes equations.\cite{GurarieRadzihovsky,LevinsenPRA} 
In this section, we discuss the solutions of this equation, focusing on
the properties of the subgap states for $\bar\mu\lesssim 1$. We first point out some general features of the analytic solution, and then deploy the formulation of 
the Bogoliubov de Gennes\cite{deGennes66} equations on the sphere,\cite{KrausPRL,KrausPRB} as well as its numerical solutions.

\subsection{Asymptotic solution of the BdG equations}
\label{sec:asymptotics}

As an approach to discussing the solutions of the BdG equations for a cylindrical vortex (\ref{eq:BdGVortex}), let us first 
\begin{figure}[tphb]
\begin{center}
 \includegraphics[width=0.8\columnwidth]{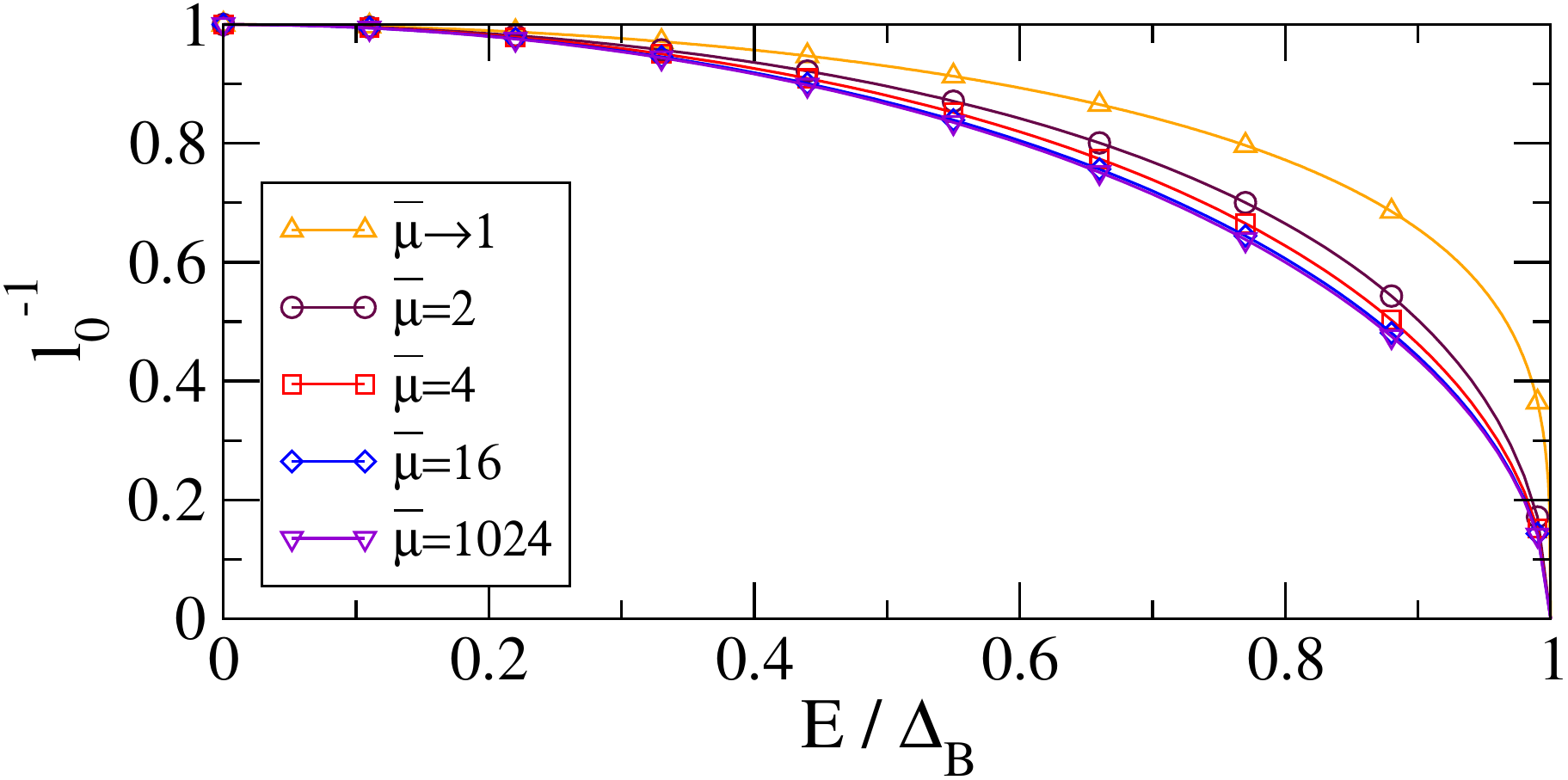}
\caption{(color online) Inverse localization length $\ell_0^{-1}$ as a function of the energy of a core state.}
\label{fig:localization}
\end{center}
\end{figure}
analyze the asymptotic behaviour in the limit of $r\to\infty$. The solutions are known to be strongly oscillatory functions, so the derivatives of the Bogoliubov 
functions are of the order of the functions themselves. However, terms in $1/r$ can be neglected, as well as $h'(r)=0$ far outside the vortex. 
Thus, (\ref{eq:BdGVortex}) reduces to
\begin{align}
\label{eq:BdGVortexAsymptotic}
-\frac{1}{2} \derivXN{r}{2}u - \bar\mu u + \derivX{r} v = \bar E u \nonumber\\
- \derivX{r} u  + \frac{1}{2} \derivXN{r}{2}v + \bar\mu v= \bar E v.
 \end{align}
Using the Ansatz of $(u\;v)^T = (A\;B)^T \exp\{i\gamma r\}$, and solving the characteristic equation for $\gamma$ at a given value of the energy yields
four possible solutions of the form
\begin{equation}
\gamma_{\pm,\pm}=\pm \left[2(\bar\mu-1) \pm 2 i \sqrt{\bar \Delta_B^2-\bar E^2}\right]^\frac{1}{2}.
\end{equation}
First, note that in the limit of $\bar E\to 0$, $\bar\mu\gg1$, these simplify to $\gamma_{\pm,\pm}\sim \pm\sqrt{2\bar\mu} \mp i$. Among the two solutions which are finite
as $r\to\infty$, we recover the behaviour of the modes (\ref{eq:zero_mode}), namely oscillations with wavenumber $\sim k_F$ and an exponential decay
with a characteristic lengthscale of localization $\ell_0=1$. In the general case, let us decompose $\gamma=k(\bar E)+i\ell_0^{-1}(\bar E)$, which, for $\bar\mu>1$ can be
written as
\begin{align}
k(\bar E)&=\pm \sqrt{2\sqrt{\bar\mu^2-\bar E^2}} \cos\left[  \frac{1}{2} \arctan\frac{\sqrt{\bar \Delta_B^2-\bar E^2}}{\bar\mu - 1} \right]\nonumber\\
\ell_0^{-1}(\bar E)&=\phantom{\pm}\sqrt{2\sqrt{\bar\mu^2-\bar E^2}} \sin\left[  \frac{1}{2} \arctan\frac{\sqrt{\bar \Delta_B^2-\bar E^2}}{\bar\mu - 1} \right].
\end{align}
According to these equations, the wavenumber of the vortex state depends only very weakly on the energy, varying between $k(0)=\sqrt{2\bar\mu-1}$ and $k(\bar \Delta_B)=\sqrt{2\bar\mu-2}$. This effect is significant only if $\bar\mu$ is small. On the other hand, the localization length is one at zero energy $\ell_0(0)=1$ and diverges as $E\to\Delta_B$. The dependency of the localization length on the energy of the subgap state is illustrated in Fig.~\ref{fig:localization} for several $\bar\mu$. For $\bar\mu$ large, the localization length goes as $\ell_0=[1-(\bar E/\bar \Delta_B)^2]^{-1/2}$.

\subsection{BdG equations on the sphere}

To study the physics of vortices in a finite size system, a convenient choice is to place a vortex-antivortex pair on the
surface of a sphere. The BdG equations for this configuration were recently described by Kraus \emph{et al}.\cite{KrausPRL} 
They are obtained by expanding the $p$-wave pairing function on the sphere 
$\Delta(\Omega, \Omega')$ in terms of the spherical monopole harmonics\cite{WuYangNuclB,WuYangPRD} $Y^q_{l,m}$, 
which also serve for the expansion of the Bogoliubov eigenfunctions $u$, $v$. For convenience, we provide the resulting
equations in dimensionless units in Appendix \ref{sec:BdGSphere}.
The relevant dimensionless parameters are given by the radius of the sphere $\bar R$, and the size of the vortex core $\bar\xi$; 
both are measured in units of the length scale $L$ from Eq.~\ref{eq:rescalings}.
The dimensionless chemical potential $\bar\mu$ is set by the Fermi angular momentum $l_F$ (number of shells filled), and also depends on the radius of the sphere
\begin{equation}
  \label{eq:chemical_potential}
  \bar\mu = \frac{l_F(l_F+1)}{2\bar R^2}.
\end{equation}

\subsection{Numerical results} 

In this section, we discuss the spectrum of Bogoliubov quasiparticle excitations on the sphere with a vortex-antivortex pair, which is obtained by solving the eigenvalue problem in its matrix form (\ref{eq:BdGSphere}) with standard linear algebra packages.

\subsubsection{Global Spectrum}
\label{Spectra}
\begin{figure}[tphb]
\begin{center}
\includegraphics[width=0.8\columnwidth]{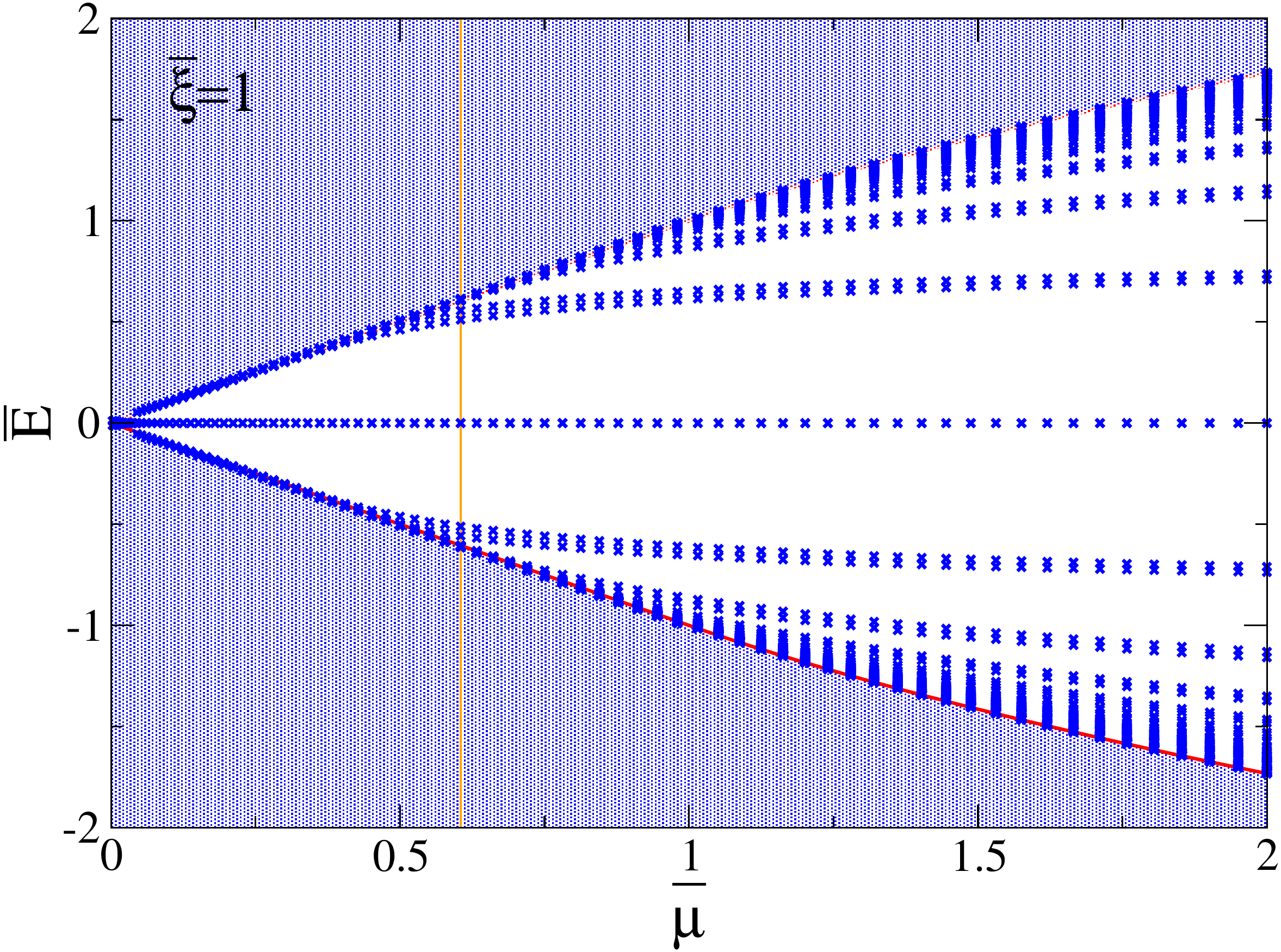}\\
\includegraphics[width=0.8\columnwidth]{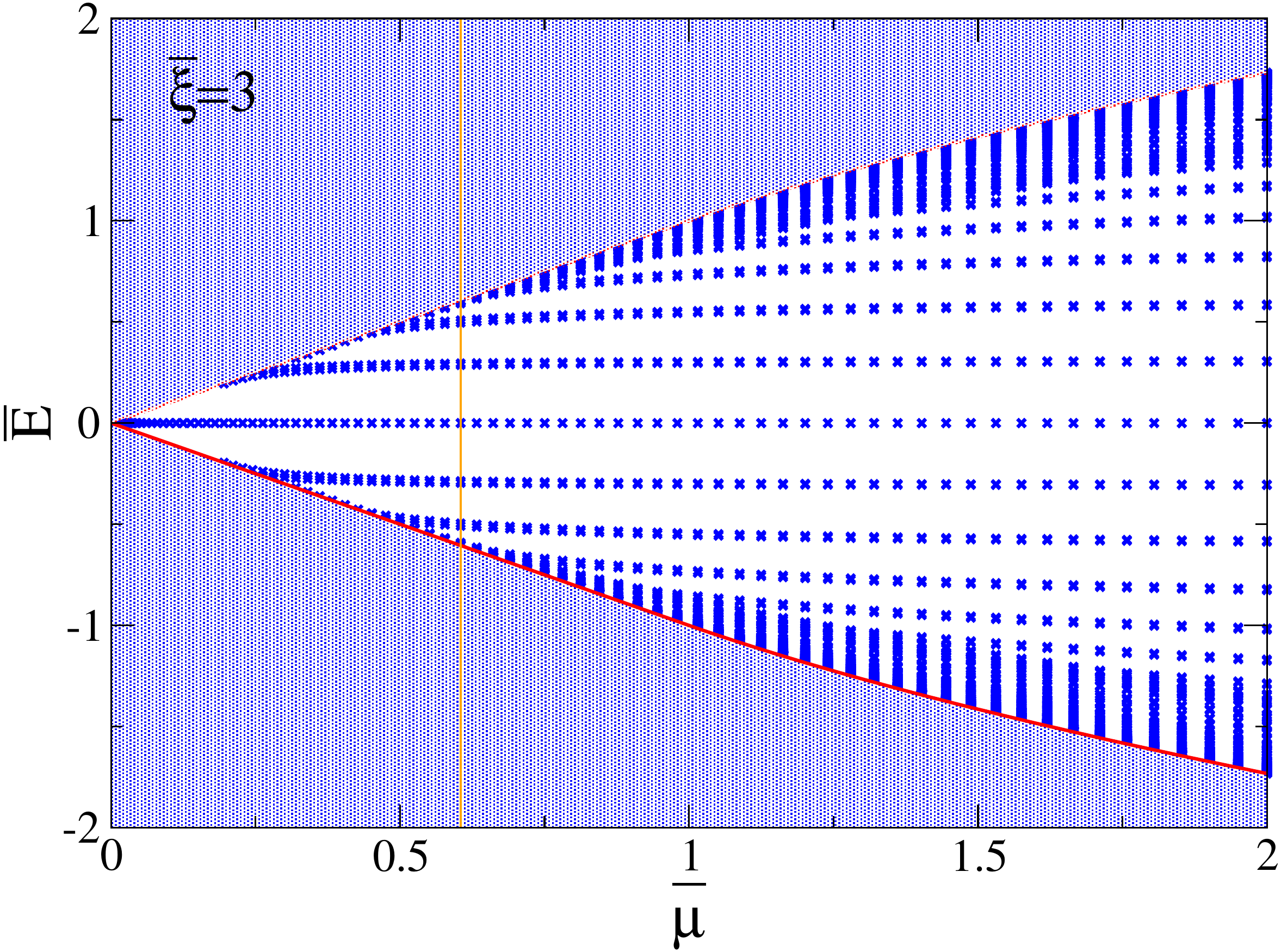}\\
\includegraphics[width=0.8\columnwidth]{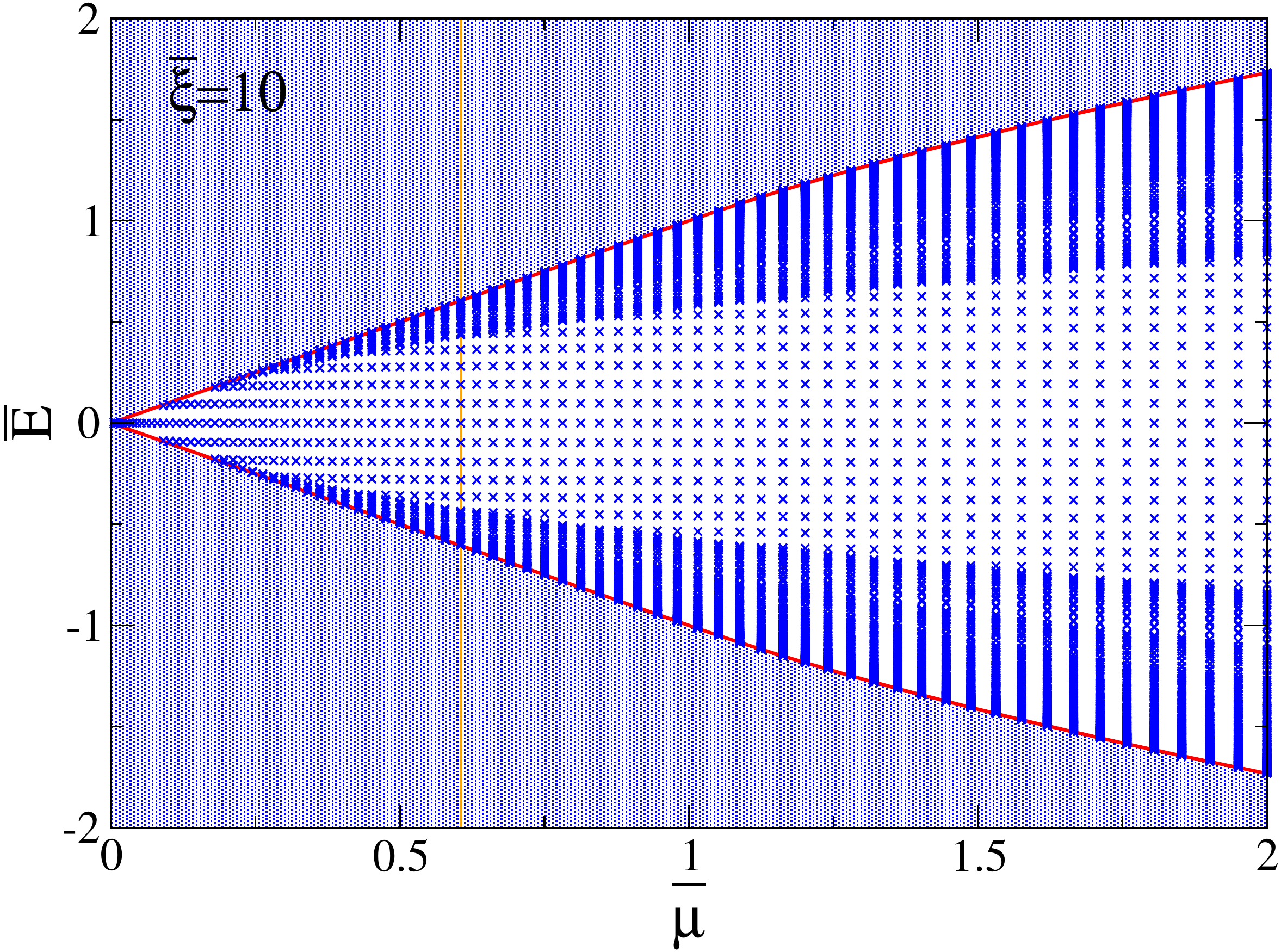}
\caption{(color online) Spectra of a $p$-wave superconductor with a vortex-antivortex pair at the antipodes of a sphere with radius $\bar R=40$ as a function of the chemical potential $\bar\mu$. From top to bottom, the figure shows the spectrum for a small vortex with $\bar\xi=1$, a vortex with $\bar\xi=3$ and a vortex with $\bar\xi=10$. Blue shaded areas indicate the presence of excitations above the gap. The subgap states are marked by crosses and include excitations below the gap up to a angular momenta with $|\langle L_z \rangle -\frac{1}{2}|<10$. We use the dimensionless units of Eq.~(\ref{eq:rescalings}). The vertical line at $\bar\mu\approx 0.6$ situates the cut through the spectrum, which is shown in Fig.~\ref{fig:spectra_m_mu_0_6} alongside $\bar\mu\approx 2$.
}
\label{fig:spectra_mu}
\end{center}
\end{figure}

 In Fig.~\ref{fig:spectra_mu}, we display the energies of eigenstates found below the bulk gap $\Delta_B$ as a function of the chemical potential, while areas above the bulk gap are shaded blue. These spectra were calculated for a sphere of radius $\bar R=40$, and include values of the Fermi angular momentum $l_F=\frac{1}{2},\ldots,\frac{159}{2}$, with a cut-off for the equation set at $l_\text{max}=200$. The bulk spectrum opens like a cone near $\bar\mu=1$ and crosses over into a square root behaviour, following Eq.~(\ref{eq:bulk_gap}). 
 
The energy of the subgap states found to depend weakly on the chemical potential. For the Caroli-de-Gennes-Matricon (CdGM) states,\cite{Caroli1964} the prediction is that the eigenstates have strictly no dependency on $\bar\mu$, as well as a linear dispersion in angular momentum [see Eq.~(\ref{eq:subgap_energy})]. Our results confirm that this is true for large enough chemical potential $\bar\mu$ and vortex size $\bar\xi$. In the bottom panel of Fig.~\ref{fig:spectra_mu}, the energy of the first subgap state $\epsilon_1$ remains roughly constant from $\bar\mu=2$ down to the point where it is absorbed into the bulk. For vortices with smaller cores, as shown in the top and centre panel of that figure, there is some `bending' of the subgap states: the energy of the subgap states goes towards a constant only for large $\bar\mu$, while at smaller chemical potential the absolute value of their energy decreases, smoothing into the cone of the propagating Bogoliubov quasiparticles. However, the energy where the subgap state is absorbed into the bulk spectrum is still of the same order as the asymptotic value. 

In addition to this slight $\bar\mu$ dependency, another feature is apparent in the spectra: while low-lying subgap states are evenly spaced, the density of states exhibits a step increase at a value below the gap, signalling the presence of additional subgap states. 

Another phenomenon, best visible in the top panel with $\bar\xi=1$, is the splitting of the subgap states. Each of the subgap states occurs as a doublet consisting of one state localized in either pair of the vortex and antivortex present on the sphere. While the splitting of the zero-modes results from the hybridization of two modes found precisely at zero energy, the splitting of the modes at non-zero energy is predominantly of a different nature: as we show in more detail below, the wavefunctions in the vortex-core are distinct for the vortex and antivortex-state for $E\neq 0$ (the splitting of the doublet saddling $E=0$ is not visible in this figure).

\subsubsection{Spectra at fixed $\bar\mu$}
\label{Dispersion}

\begin{figure*}[ptbh]
\begin{center}
\includegraphics[width=0.8\columnwidth]{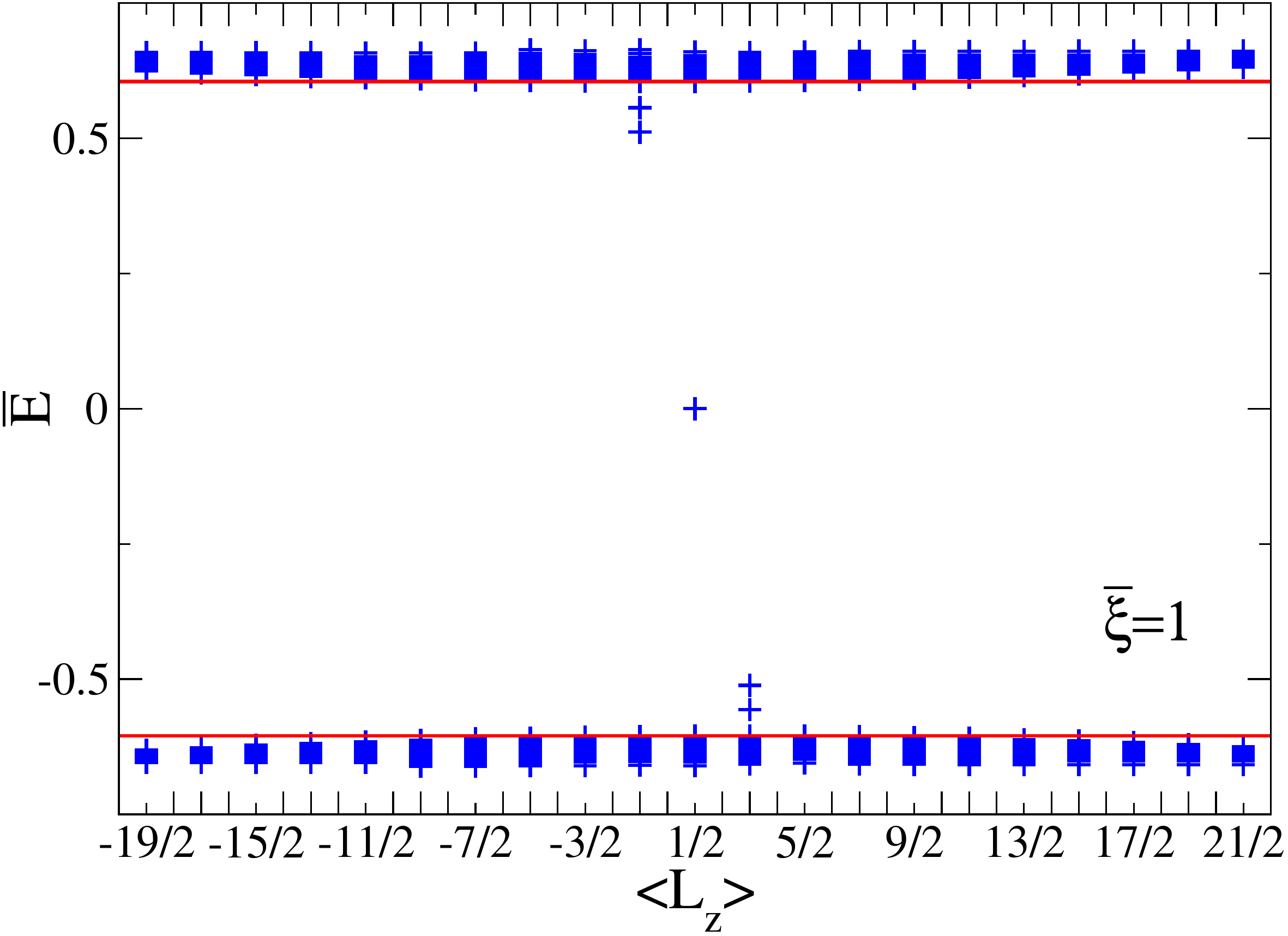} $\quad$ \includegraphics[width=0.8\columnwidth]{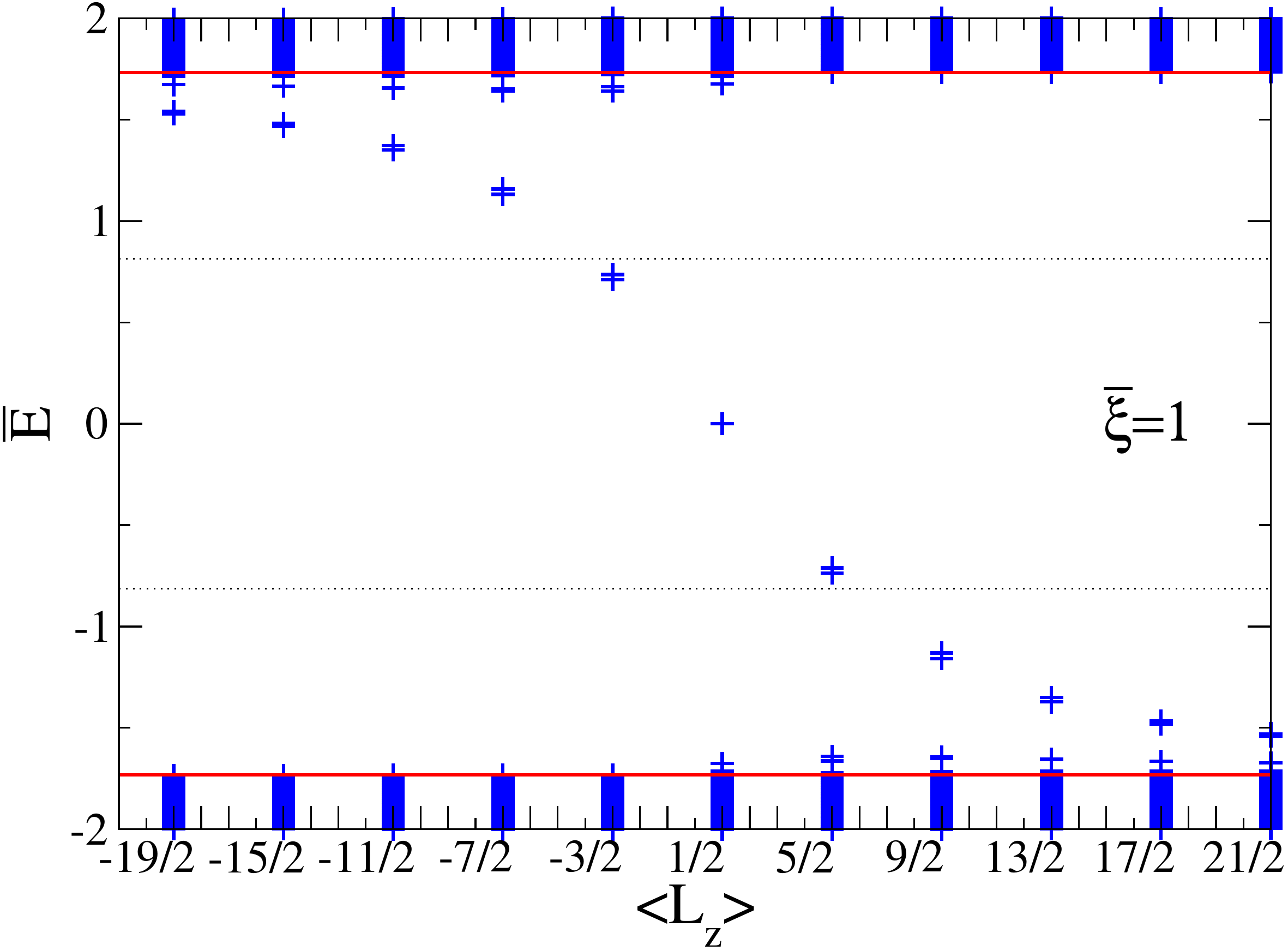}\\
\includegraphics[width=0.8\columnwidth]{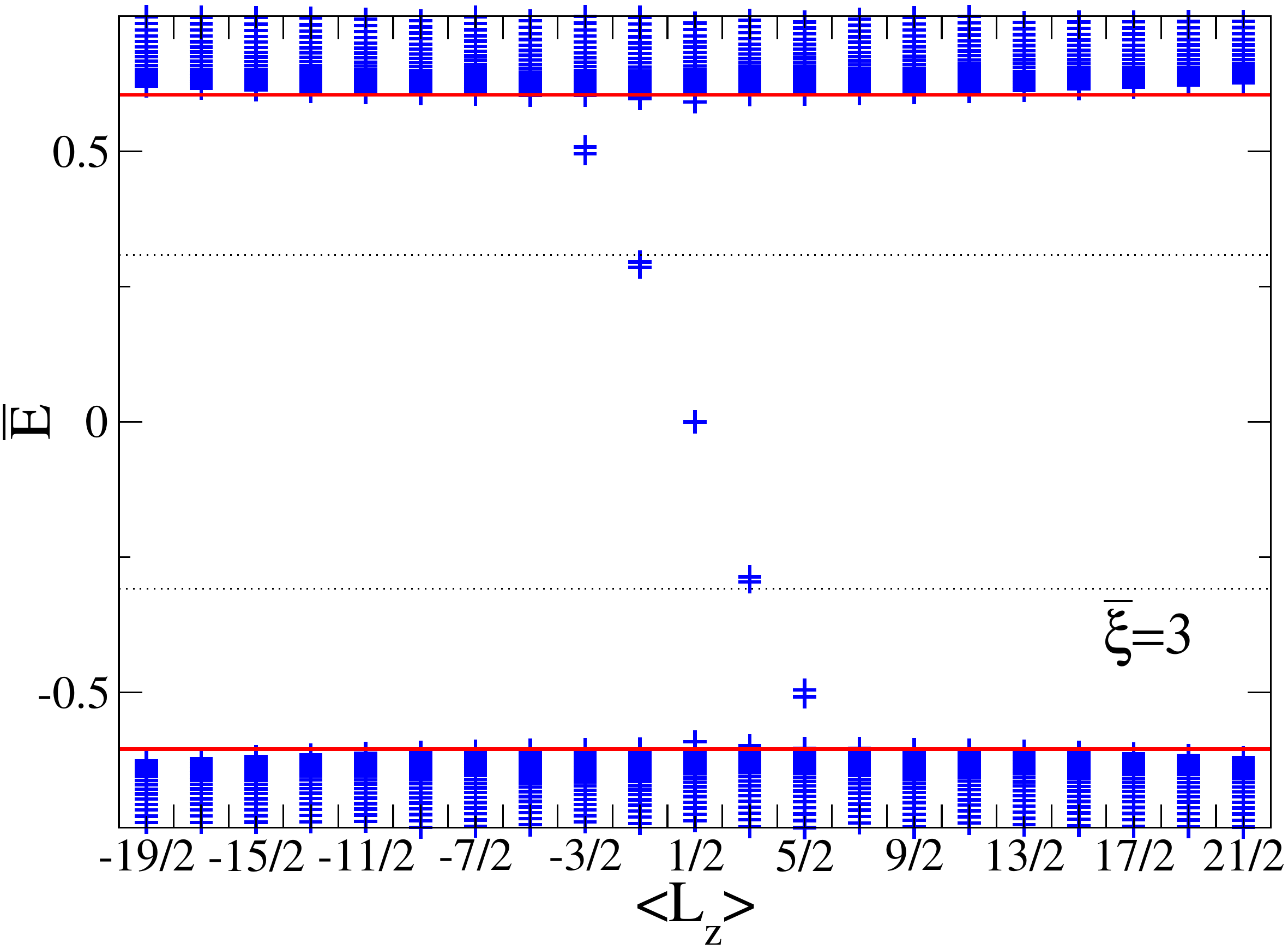} $\quad$ \includegraphics[width=0.8\columnwidth]{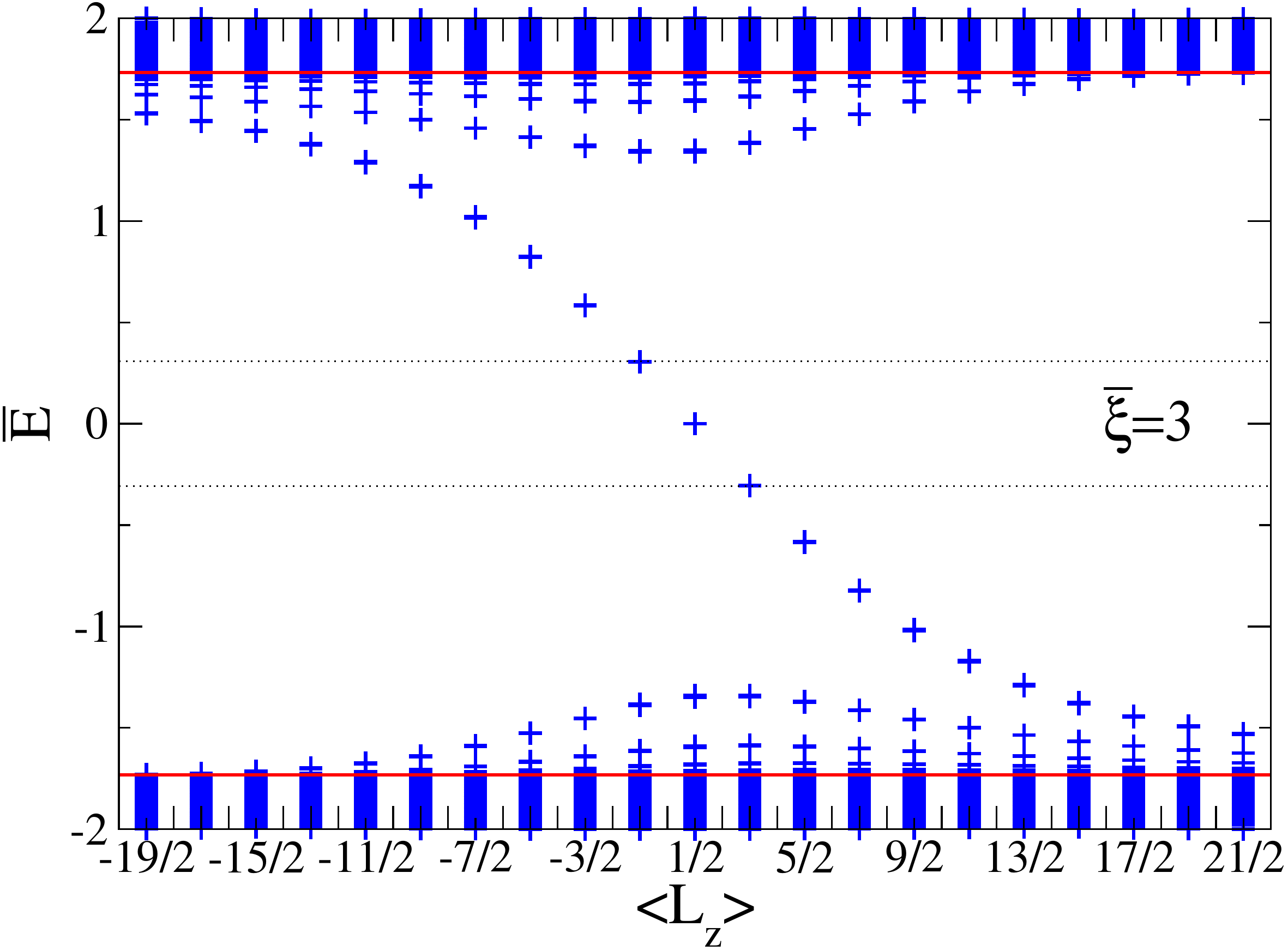}\\
\includegraphics[width=0.8\columnwidth]{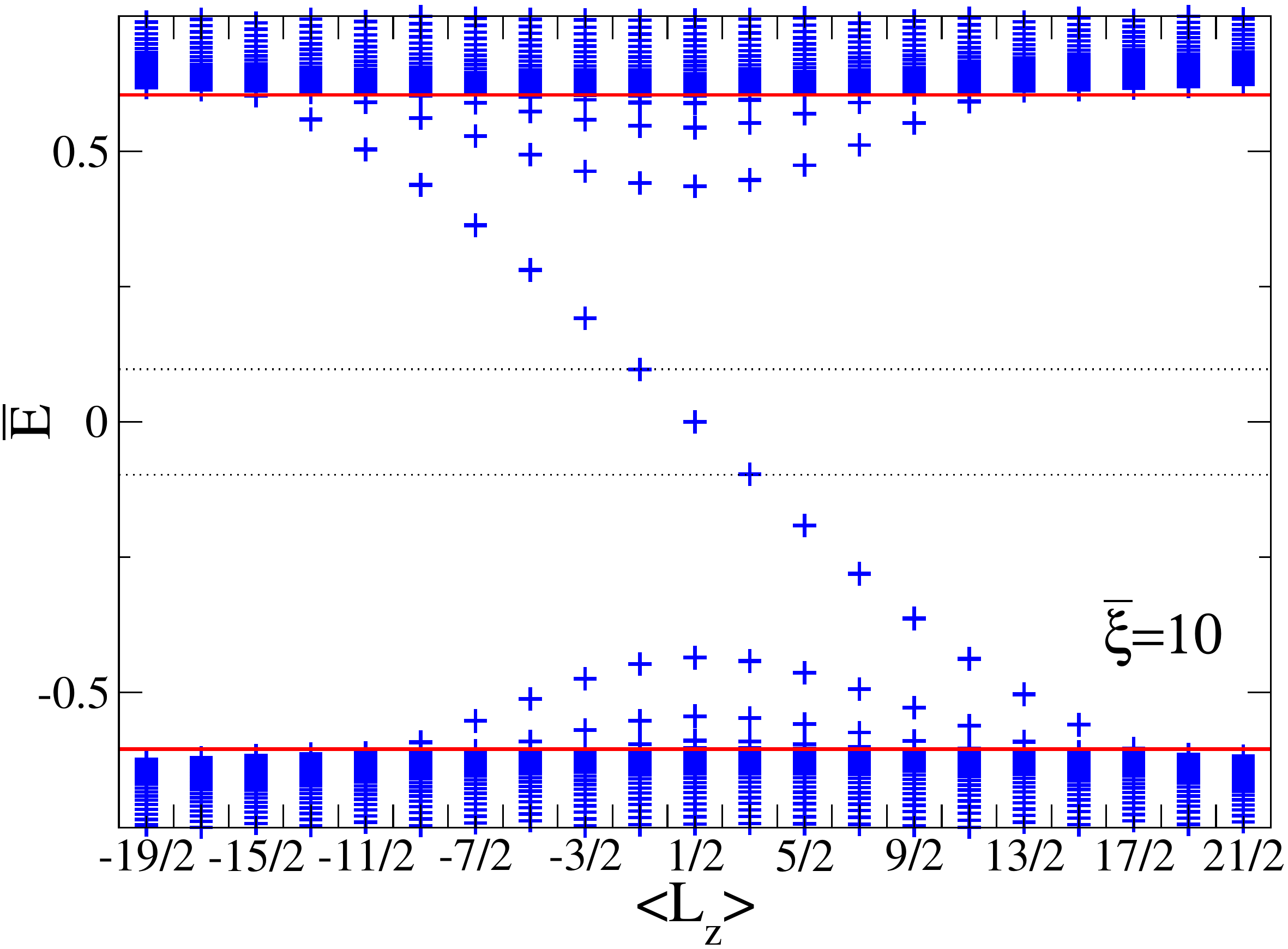} $\quad$ \includegraphics[width=0.8\columnwidth]{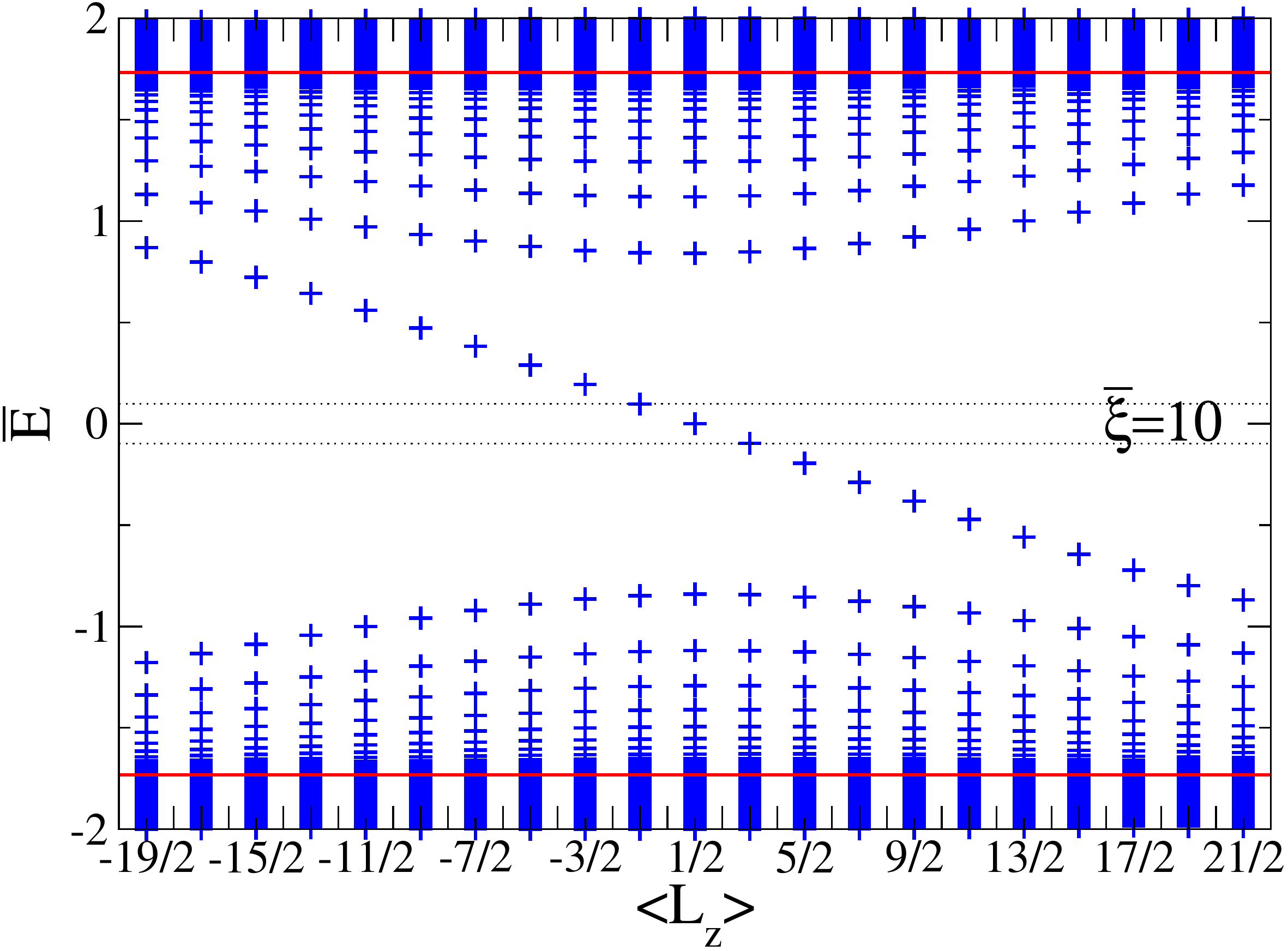}
\caption{(color online) Spectra as a function of the angular momentum $m=\langle L_z \rangle$ for a $p$-wave 
superconductor with a vortex-antivortex pair at the antipodes of a 
sphere with radius $\bar R=40$ at fixed chemical potential $\bar\mu$. The left column shows $\bar\mu\approx 0.6$ (according to the yellow cuts in Fig.~\ref{fig:spectra_mu}), and the right column shows $\bar\mu=2$. Vertically, we show configurations for a small (top, $\bar\xi=1$), medium (centre, $\bar\xi=3$) and large (bottom, $\bar\xi=10$) vortex respectively, as in Fig.~\ref{fig:spectra_mu}. The expected bulk gap $\bar\Delta_B$ is shown as a solid red line, the energy $\bar\omega_0$ according to (\ref{eq:subgap_spacing}) as dotted black lines. The subgap states can be classified as two types: the anomalous branch of the CdGM states are approximately linearly (for small $m$) dispersing near $E=0$, with a negative slope. Additional subgap modes with radial quantum number $n\neq 0$ occur in large vortices and have with a finite minimum (maximum) of $|\bar E|\sim\mathcal{O}(\bar \Delta_B)$ at/near $m=\frac{1}{2}$.  The number $n\neq 0$ modes  increases with $\bar\mu$.}
\label{fig:spectra_m_mu_0_6}
\end{center}
\end{figure*}

The dispersion of the spectra as a function of the angular momentum $m=\langle L_z \rangle $ relative to the symmetry axis of the vortex cores reveals some additional insights. For simplicity, the discussion focuses on the states with $E>0$, keeping in mind that states at negative energy are related by virtue of the symmetry of the Hamiltonian $E(m)=-E(1-m)$. We display the dispersion at two distinct values of the chemical potential. Figure \ref{fig:spectra_m_mu_0_6} shows the energy as a function of $m$ for $\bar\mu\approx 0.6$ (left column) and $\bar\mu\approx 2$ (right column) respectively, in the geometries already used for Fig.~\ref{fig:spectra_mu} (corresponding to values of $l_F=\frac{87}{2}$ and $l_F=\frac{159}{2}$). 

Let us first discuss the case of $\bar\mu=0.6$. Irrespective of the size of the vortex core, we find that the bulk gap $\bar\Delta_B$ (solid red lines) conforms well with the case of a homogeneous order parameter without vortices, as given by Eq.~(\ref{eq:bulk_gap}).  By contrast the estimate for the energy of the first subgap state (black dotted lines), obtained from (\ref{eq:subgap_spacing}) using the same radial profile as in (\ref{eq:vortex_field}), is much less accurate, in particular for the small vortex cores. For $\bar\xi=1$, the predicted subgap energy $\bar \omega_0(\bar\xi=1)\approx 0.814$ is roughly $1.5$-fold larger than the actual eigenstates $\bar \epsilon_1$. The chemical potential here was chosen such that the first subgap state almost merges with the bulk spectrum. The corresponding pair has a very large splitting of $\delta_1=0.042$, almost 8\% of their median eigenvalue $\epsilon_1$. The splitting of the states is analysed in more detail, below. 
With increasing size of the vortex core, the previous estimate of $\epsilon_1\approx\omega_0$ from Eq.~(\ref{eq:subgap_spacing}) is increasingly accurate. However, in this regime of small $\bar\mu$, the subgap states have a dispersion which is sub-linear, i.e., the mode of the CdGM states bends to asymptote the bulk gap. 

In the bottom panel of Figure \ref{fig:spectra_m_mu_0_6}, it is apparent that there can be multiple subgap states at a given angular momentum $m$. Like the CdGM states, these modes consist of pairs of eigenstates at each value of $m$, as expected for localized states in the presence of two vortex cores. We refer to these states by an additional quantum number $n$, where $n=0$ refers to the CdGM branch. In addition, we denote the two degenerate states of each twofold degenerate branch by $+/-$. The dispersion of the $n\neq 0$ modes has a minimum in $|E|$ at or near $m=\frac{1}{2}$. On first sight there appears to be a symmetry relating eigenstates of $E(m)\leftrightarrow E(1-m)$, but the spectrum is slightly skewed such that $E(\frac{1}{2}+\delta m)>E(\frac{1}{2}-\delta m)$, given ($\delta m=1,2,\ldots > 0$). This is unlike the case of $s$-wave superconductors where  the $n\neq 0$ modes are characterized by a symmetry of the spectrum in $E(m)=E(-m)$.\cite{KopninLopatin95}  For large $|m|$, the modes asymptote towards the bulk gap. Near their minimum, the dispersion of the $n\neq 0$ modes is roughly quadratic, leading to the markedly higher density of states as compared to the $n=0$ CdGM states, and thus explaining the jump in the density of states that we pointed out in Fig.~\ref{fig:spectra_mu}.

We are not aware of other numerical studies that have analyzed the $n\neq 0$ subgap states, as in typical type II superconductors, the vortex core is too small for such additional bound states to exist.\cite{GygiSchluter91} Within BCS theory, we found above in section \ref{sec:dipolar} that the coherence length of the superfluid is generically near one. In atomic $p$-wave superfluids, the existence of the dimensionless parameter $c_2$ allows one to scale the coherence length, which may become large for small $c_2$ according to Eq.~(\ref{eq:coherence}). Therefore, discuss their properties in more detail in section \ref{sec:branches}, below.

Due to the asymmetry of the $n\neq 0$ subgap modes between positive and negative angular momenta, they may influence the dynamics of the vortices
in superfluids. In particular, the Magnus-force acting on vortices that are moving through the system may acquire corrections resulting from the influence of the
subgap states,\cite{VolovikFlow93A, VolovikFlow93B, KopninSalomaa,KopninLopatin95,StoneReview} although it was debated whether such influences should vanish as the Magnus force can be derived from a  Berry phase.\cite{AoThouless} In $s$-wave superconductors, the CdGM states would be singled out as the only mode contributing to corrections to the Magnus force, as these corrections vanish if $\derivX{k}E_n(k)$ is odd.\cite{KopninLopatin95} In $p$-wave superfluids, $\derivX{k}E_n(k)$ does not have this symmetry and thus, $n\neq 0$ modes may contribute. It is difficult to estimate the magnitude of the additional corrections: while the asymmetry of $\derivX{k}E_n(k)$ is small, the density of states contributing to corrections is much higher than for the CdGM mode.

\subsubsection{Splitting of the CdGM states}

In section \ref{Spectra} above, we noted that the splitting of the CdGM states in the spectra for the vortex-antivortex pair is much larger at 
$E\neq 0$ than for the Majorana zero-modes. For the latter, this splitting is interpreted as arising from a tunnelling term that hybridizes the 
two degenerate modes. 
\begin{figure}[ttbp]
\begin{center}
\includegraphics[width=0.95\columnwidth]{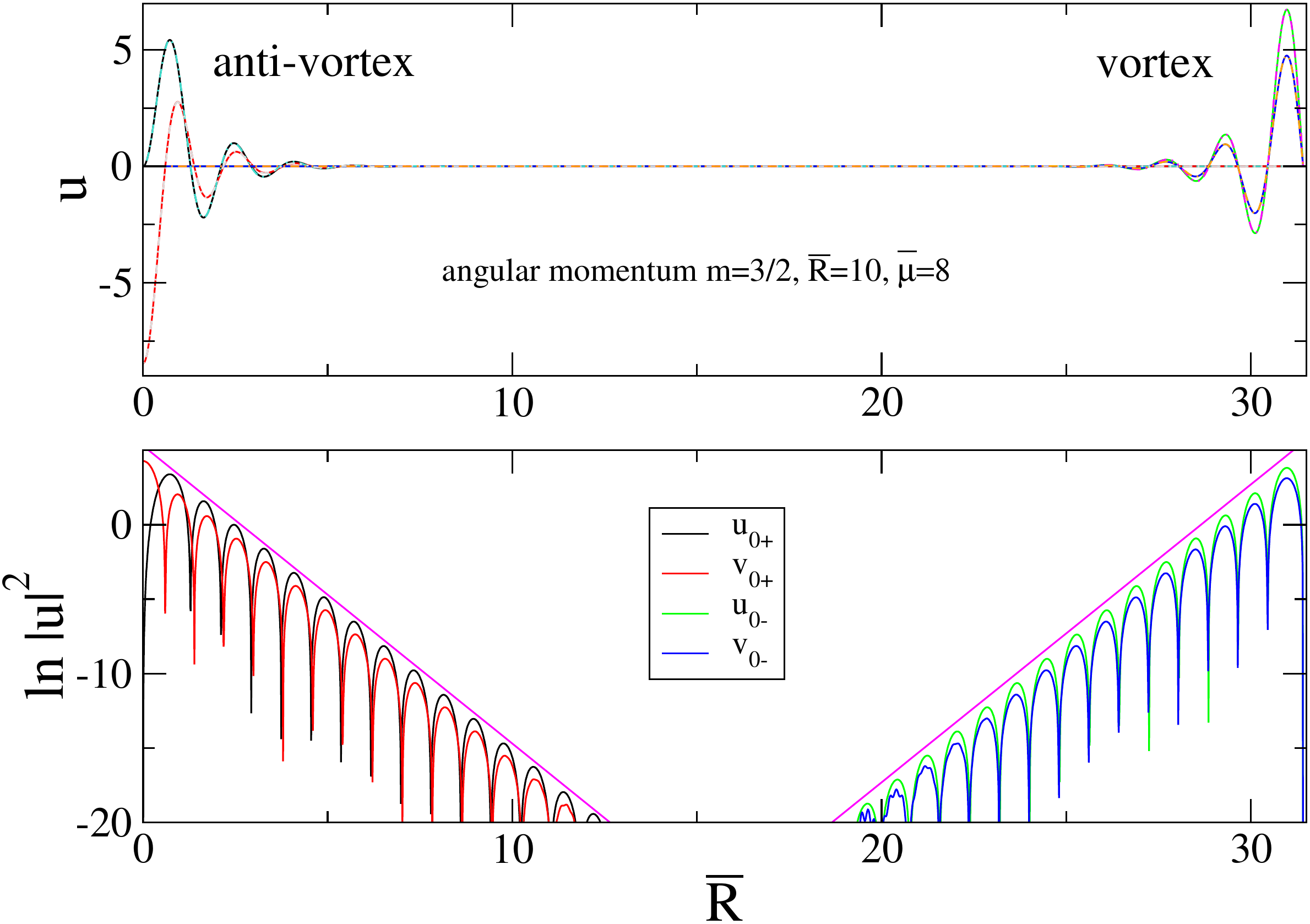}
\caption{(color online) The wavefunctions $u(r)$, $v(r)$ for the first excited state of the CdGM branch with $n=0$ and $m=3/2$, for a sphere of radius $\bar R=10$ and chemical potential $\bar\mu=8$. The eigenvalues associated with the upper and lower eigenvalue are split and amount to $\bar\epsilon_{0+}=-0.441639$ and $\bar\epsilon_{0-}=-0.443256$, respectively. The upper panel shows $u$ and $v$ superposed with a fit that shows excellent agreement for the behaviour in the vortex core. The lower panel displays the wavefunctions on a logarithmic scale and indicates the expected exponential decay for comparison.}
\label{fig:wavefunctions} 
\end{center}
\end{figure}

\begin{figure}[tbhp]
\begin{center}
\includegraphics[width=0.95\columnwidth]{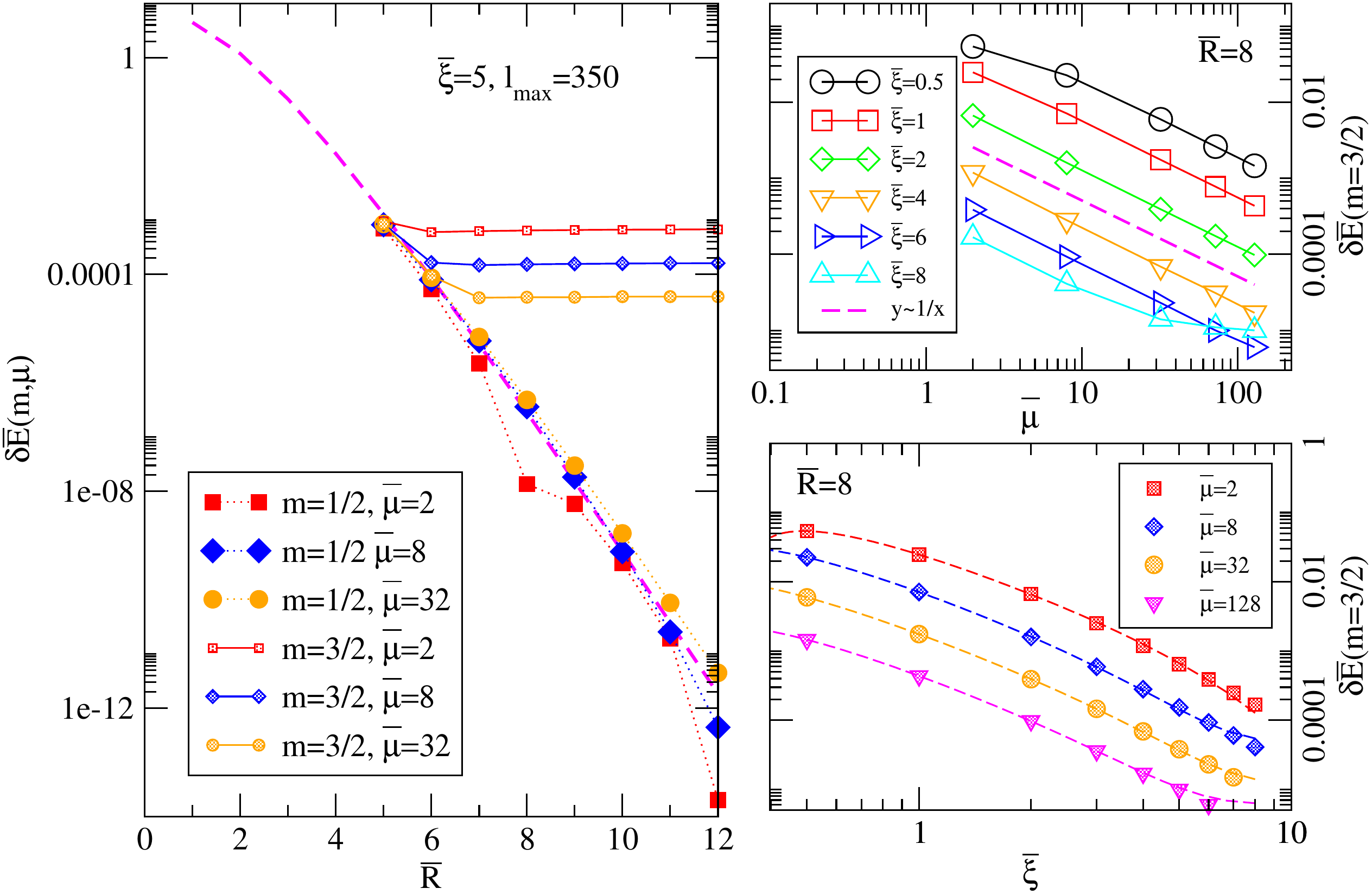}
\caption{(color online) Left panel: The splitting of the zero modes and first CdGM subgap states at $m=3/2$ is shown as a function of the radius $\bar R$ of the sphere geometry for several values of $\bar\mu$.
The splitting of the zero modes follows the amplitude of the envelope of the wavefunction at $\bar R/2$ (magenta dashed). For small $\bar\mu$, some oscillations are seen, as displayed separately below in Fig.~\ref{fig:splitting_mu_small}. At large enough vortex separation $r=\pi \bar R$, the splitting of the $m=3/2$ states is given by the shape of the vortex core and independent of $r$. At small $r$ it is increased by the hybridization of the modes. Top right panel: splitting of the subgap states $\delta \bar E$ at $m=3/2$ as a function of $\bar\mu$, found to be inversely proportional. Bottom right panel:  the same splitting $\delta \bar E(m=3/2)$ as a function of the vortex size. The dashed lines show a fit with a third order polynomial in $\bar\xi^{-1}$.}
\label{fig:splitting_detail}
\end{center}
\end{figure}

Contrary to the case of the zero modes, subgap states at $m\neq \frac{1}{2}$ (or $E\neq 0$) are not precisely degenerate even for a well separated 
vortex-antivortex pair. This may be suprising at first, as within the perturbative solution\cite{KopninSalomaa} of the BdG equations, the energy of the 
subgap states does not depend on the vorticity. 

However, the BdG equations for different vorticity $\kappa$ are distinct: as Eq.~(\ref{eq:BdGVortex}) shows, the equations differ in 
the index of the differential operator $\mathcal{D}_l$, resulting in different short distance behaviour via the terms $[m\pm(\kappa-1)/2]/r^2$. 
Indeed, the perturbative solution yields wavefunctions proportional to the Bessel functions $u_{m,\kappa}(r) \sim J_{m+\frac{\kappa-1}{2}}(r)$ 
and $v_{m,\kappa}(r) \sim J_{m-\frac{\kappa-1}{2}}(r)$ at small $r$. According to this solution, $u$ and $v$ have the same short distance 
behaviour behaviour in a vortex, while they are proportional to two distinct Bessel-functions in the antivortex, with an index differing by two. 
(Only for the zero-mode this offset is trivial as it pairs $u\sim J_1$ and $v\sim J_{-1}$, which are equal up to a sign.) It then becomes obvious
 that the energy of the eigenstates depends on the detailed interaction between the shape of wavefunctions and the order parameter within 
 the vortex-core that is neglected in the perturbative solution. For illustration, Fig.~\ref{fig:wavefunctions} displays the eigenfunctions for the first 
 subgap state, showing clearly the distinction between the cases of a vortex and antivortex.

A detailed analysis of this situation is presented in Fig.~\ref{fig:splitting_detail}, where we focus in particular on the splitting of the zero mode, as 
well as the first subgap state. In the left panel, the magnitude of these splittings is plotted as a function of the radius of the underlying sphere $\bar R$,
which sets the distance of the vortex-antivortex pair. The splitting of the zero mode is described very well by the square of the envelope of the 
wavefunctions at half the sphere radius $\exp[-2\int_0^{\bar R/2} h^2(r)dr]\approx \exp[-2 \bar R]$, for low-lying subgap states. We discuss some 
deviations observed at small chemical potential in more detail, below. For the first excited state, the splitting falls onto this curve only if the radius 
is very small. At large separation, it saturates as a constant value and is entirely set by the physics inside the vortex core, as argued in the previous 
paragraph. The two panels on the right-hand side indicate the magnitude of the splitting for the first subgap state as a function of either the chemical 
potential $\bar\mu$ and the vortex size $\bar\xi$. The functional dependency goes as $\delta \bar E(m=3/2) \sim \bar\mu^{-1}$. Its relationship to the size of the vortex 
core is less clear, but it can be fit well by a third order polynomial in the inverse core size, in accordance with the notion that the shape of the vortex 
core strongly influences this energy scale.

\begin{figure}[bpth]
\begin{center}
\includegraphics[width=0.9\columnwidth]{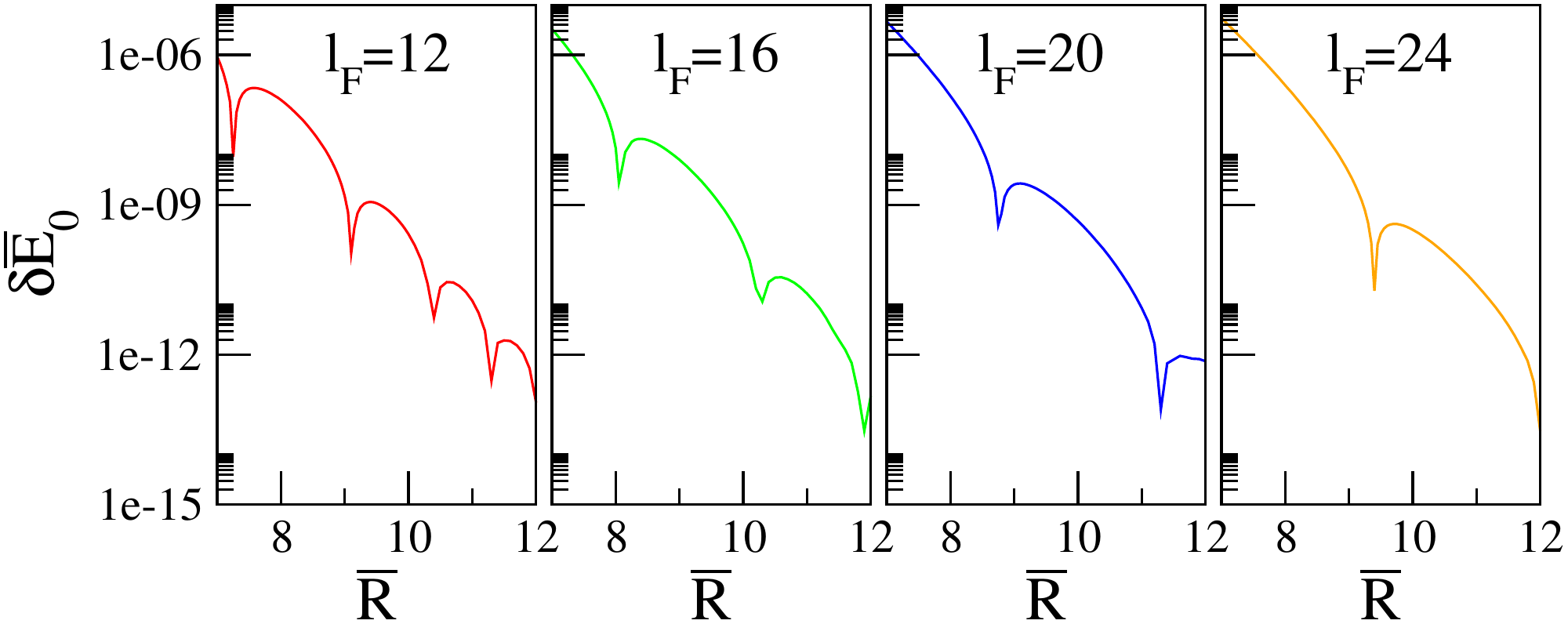}
\caption{(color online) Splitting of the zero modes $\delta E_0$ at $m=\frac{1}{2}$ for small values of the Fermi angular momentum ranging from $l_F=12$ (leftmost panel) to $l_F=24$ (rightmost panel), plotted as a function of the sphere radius $\bar R$. See main text for a discussion.}
\label{fig:splitting_mu_small}
\end{center}
\end{figure}

The amplitude for tunnelling processes between vortices is set by the overlap of the respective wavefunctions. As the eigenstates are strongly
oscillatory, the tunnelling amplitude as a function of the vortex separation is not merely set by the exponential envelope, but it additionally changes 
sign periodically in the separation between vortices, as recently discussed in the literature.\cite{BarabanSplitting09, MengSplitting09} It may be surprising
at first that no such oscillations are seen here. However, the configuration we study is very special in that the circumference of the sphere is close an
integer multiple of the Fermi-wavelength for large $l_F$ according to (\ref{eq:chemical_potential}). Deviations from this situation are thus found only 
at small values of the Fermi angular momentum. Indeed, we find that oscillations can be seen for small $l_F$, as shown in Fig.~\ref{fig:splitting_mu_small}. 
Given that the chemical potential can take only the discrete set of values (\ref{eq:chemical_potential}), we plot the splitting at constant $l_F$, rather 
than constant $\bar\mu$. Note that the oscillations we see occur at a lower frequency for larger chemical potential, even though the wavelength of oscillations 
for the wavefunctions of these states increases, in line with our explanation in terms of multiple Fermi-wavelength filling the circumference of the sphere.
The oscillating behaviour of the subgap states was recently explored numerically in the plane geometry.\cite{MizushimaMachida10b}

\subsubsection{Subgap states: branches with $n\neq 0$}
\label{sec:branches}
In large vortex cores, multiple branches of subgap states occur, which can be interpreted as states with a different radial quantum number $n$ (the CdGM being $n=0$). Here, we analyse some properties of the eigenstates at angular momentum $m=\frac{1}{2}$ at different values of $\bar\xi$ and $\bar\mu$, with eigenvalues denoted as $E_{m,n}$. 

\begin{figure}[tbbp]
\begin{center}
\includegraphics[width=0.9\columnwidth]{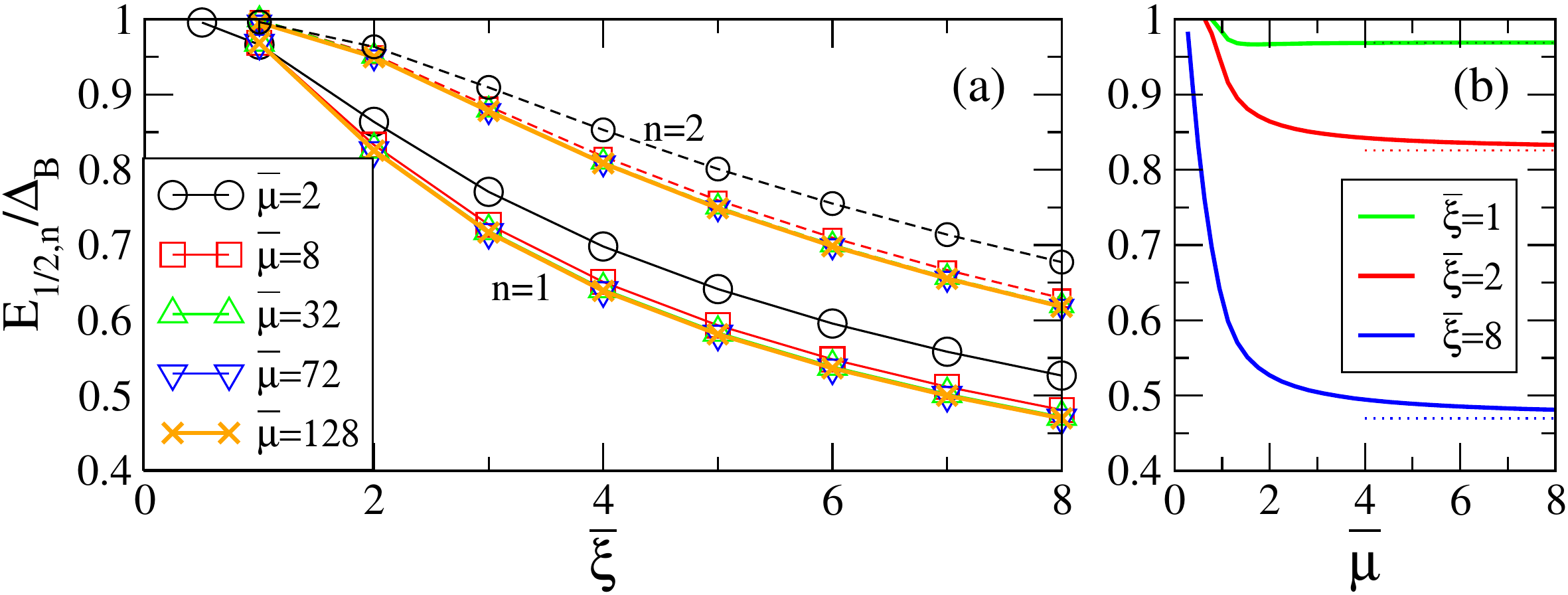}
\caption{(color online) (a) The energy $E_{1/2,1}$ of the eigenvalue at angular momentum $m=\frac{1}{2}$ for branches of subgap states with radial quantum number $n=1$ (solid lines) and $n=2$ (dashed lines), for different chemical potentials $\bar\mu$. This data was collected for a sphere with radius $\bar R=8$. When expressed in units of the bulk gap $\Delta_B$ (as shown), the dependency $E/\Delta_B$ approximately collapse onto a single curve for large $\bar\mu$. (b) Showing the dependency of $E_{1/2,1}/\Delta_B$ on the chemical potential for the $n=1$ branches of the subgap states: the additional branches are pushed into the continuum for small $\bar\mu$.}
\label{fig:branches}
\end{center}
\end{figure}

As displayed in Fig.~\ref{fig:branches}, our numerical results show that additional modes occur for vortices with size $\bar\xi\gtrsim 1$. The energy of the $n>1$ branches can be expressed in fractions of the bulk gap: to a good approximation, the eigenvalues do not depend on the chemical potential (except for small $\bar\mu$), and $\bar E_{\frac{1}{2},n} (\bar\xi,\bar\mu) = f_n(\bar\xi) \bar \Delta_B(\bar\mu)$.

\begin{figure}[tbhp]
\begin{center}
\includegraphics[width=0.95\columnwidth]{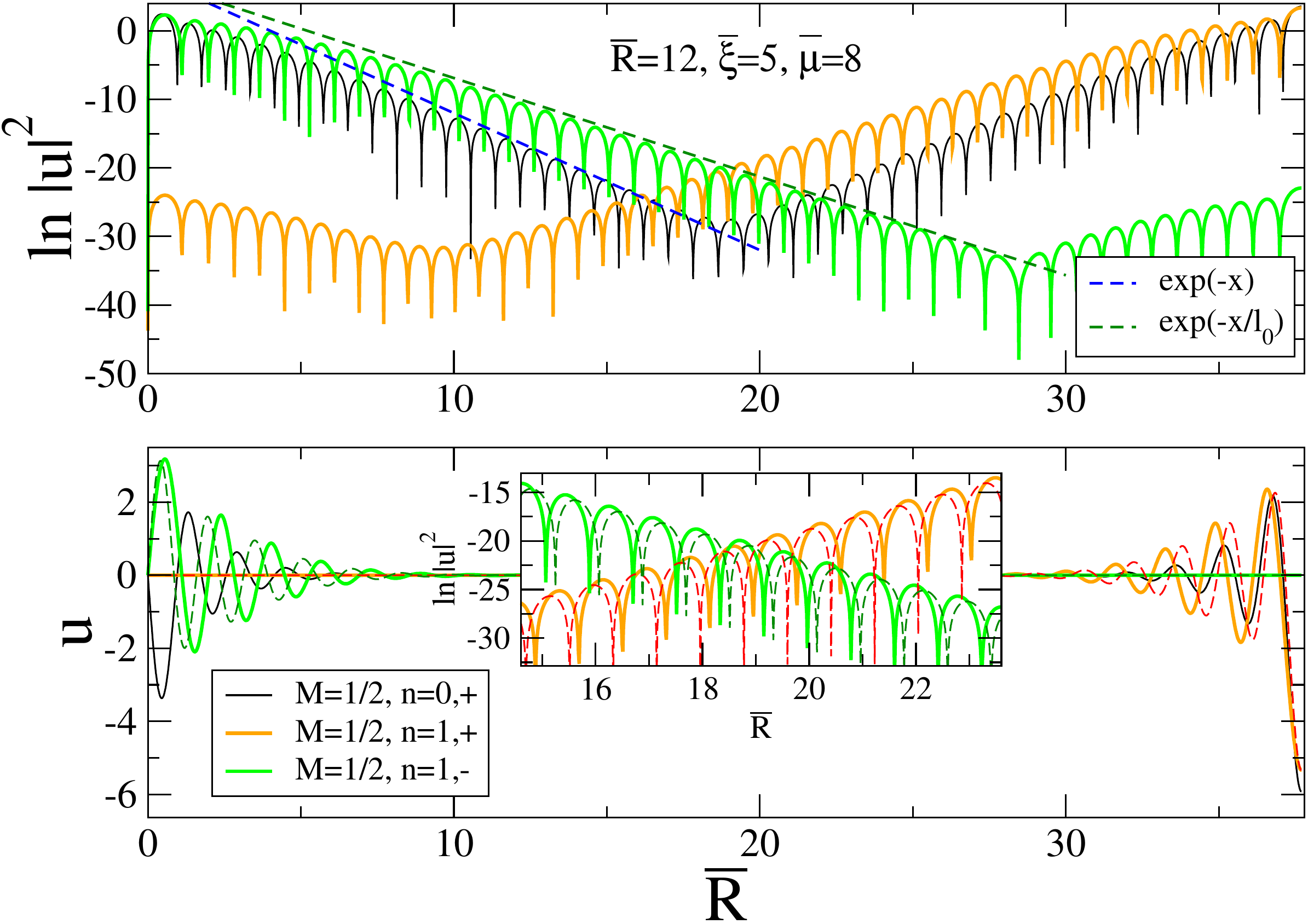}
\caption{(color online) Wavefunctions $u$ for the zero mode and a nearly degenerate pair of the $n=1$ modes at $m=1/2$. Upper panel: $u$ in logarithmic scale, showing the exponential localization of these modes, which is found to be weaker for the excited states. Note the hybridization of the zero-modes, while each of the excited states is localized near a single vortex. Lower panel: showing the Bogoliubov functions $u(r)$ and in dashed $v(r)$ for the same modes. Note the phase shift between $u$ and $v$ for the excited state. The inset, again using a logarithmic scale, shows that the phase-shift remains constant far from the vortex core.}
\label{fig:wavefunctions_n_1}
\end{center}
\end{figure}

We now discuss features of the wavefunctions of the $n>0$ modes. Firstly, as these states occur at energies which are large fractions of the bulk gap, they are less strongly localized than the low-lying subgap states. An example set of wavefunctions for a vortex pair with $\bar\xi=5$ at $\bar\mu=8$ on a sphere with $\bar R=12$ is shown in Fig.~\ref{fig:wavefunctions_n_1}, where the upper panel gives a comparison of the $n=1$ state (occurring at $E_{1/2,1}/\Delta_B\approx 0.6$) and the $n=0$ zero mode in logarithmic scale, showing the excited state with a localization length of about $\ell_0=1.35$, compared to $\ell_0=1$ for the groundstate. 
This localization length is slightly larger than the value predicted from the asymptotic solution ($\ell_0=1.23$) of the simplified set of BdG equations (\ref{eq:BdGVortexAsymptotic}) that we discussed in section \ref{sec:asymptotics}. As the size of the vortex is reduced, and the chemical potential is increased, the localization length converges to the asymptotic estimate. 

Secondly, another interesting feature of the $n>0$ modes is the occurrence of a phase shift between
the $u$ and $v$ functions, which is displayed in the lower panel of Fig.~\ref{fig:wavefunctions_n_1}.
In the vortex-core, the two functions are oscillating at slightly different wavelengths. Far from the vortex core, they have sinusoidal forms of the same decreasing amplitude, but with a constant phase shift.

\subsection{Experimental Consequences in Cold Atomic Gases}
\label{sec:exp_consequences}

To illustrate our findings in more familiar units, we devote this section to discussing the orders of magnitude of the different energy-scales 
involved in typical experiments for the known Feshbach resonance in cold $^{40}$K gases. As shown in Appendix \ref{sec:experimentalC2},
the scattering parameters which
ultimately determine the physics of the $p$-wave superfluid can be evaluated explicitly from experimental data. This analysis yields the magnitudes
of the coupling constant $g$ of the two-channel model (\ref{eq:hamil}), as well as the dimensionless constant $c_2$ which sets the maximum 
density of bosons, and thus the coherence length $\bar\xi=\sqrt{c_2/(1+c_2)}$ in our dimensionless units. 
\begin{figure}[tbhp]
\begin{center}
\includegraphics[width=0.8\columnwidth]{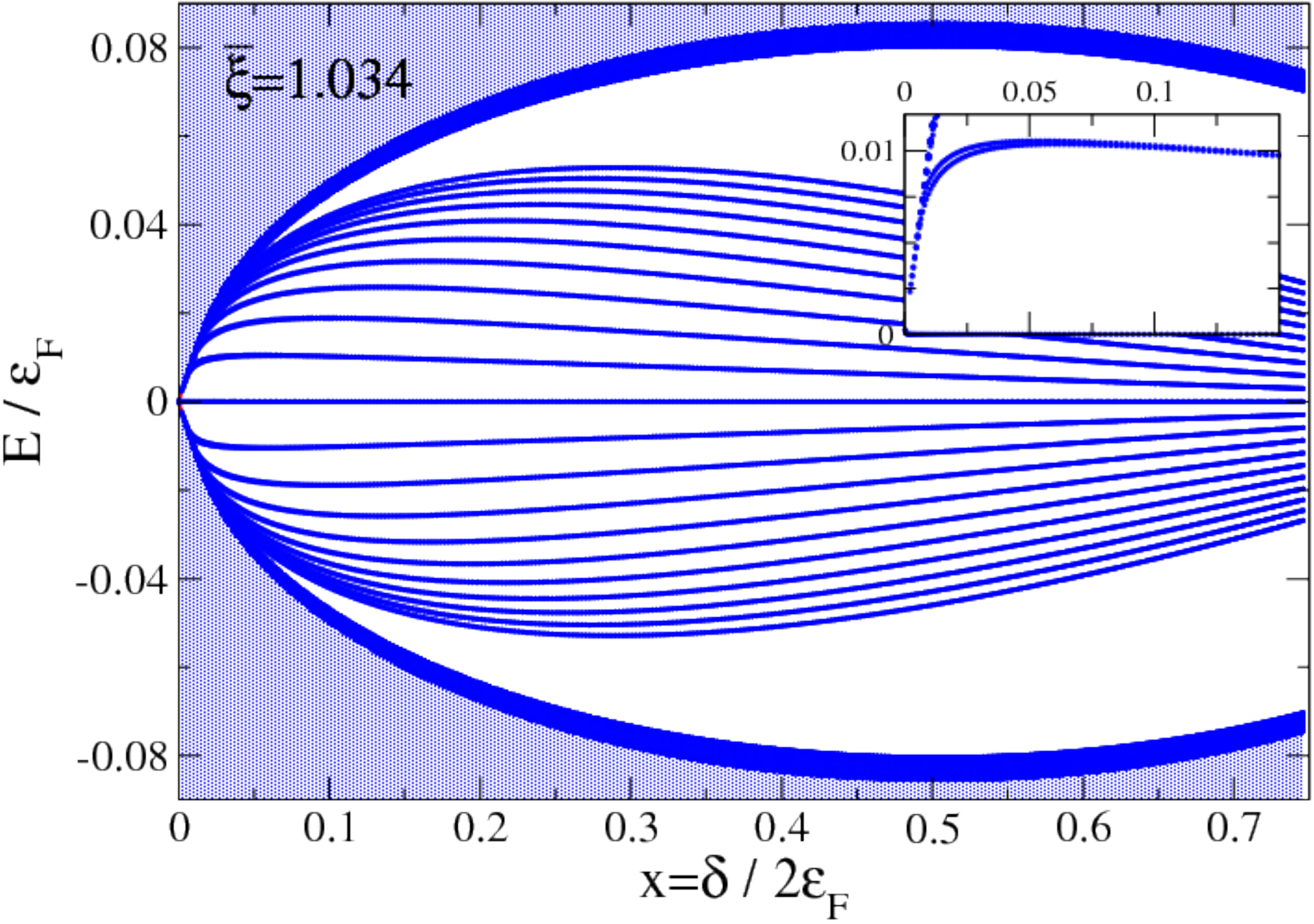}
\caption{(color online) 
Using the values thus obtained, of
$c_2\approx 14.4$ and $g\approx 1.4 \cdot 10 ^{-46}Jm^{\frac{5}{2}}$, the subgap spectrum can be given as a function of the detuning
$\delta$, resulting in Fig.~\ref{fig:spectrum_dimensions}. 
Spectrum of the subgap states, in units of Fermi energy, as a function of the detuning $\delta$ for the Feshbach resonance in a $^{40}$K gas. The
graph includes subgap states with angular momentum $|m|\leq 10$; higher angular momentum states would lie between the displayed subgap states and the bulk gap. 
The largest mini-gap is obtained at a detuning $\delta\approx 0.062$, and amounts to roughly 1\% of the Fermi energy, as highlighted in the inset.}
\label{fig:spectrum_dimensions}
\end{center}
\end{figure}
The scaling of the axes relative to the dimensionless units used in the bulk of this paper is
set by the factor $\mathcal{S}$ [see (\ref{eq:scaleS})], which amounts to $68.1$ for the specific case of $^{40}$K, further assuming a confinement length 
of $\ell=500nm$. In particular, the overall scale of energies becomes $E=\mathcal{S}^{-1}\bar E \epsilon_F$, in terms of our dimensionless 
energies $\bar E$. As $\mathcal{S}$ is only moderately large for $^{40}$K, 
the mini-gap can be as large as 1\% of the Fermi-energy. The maximum of $\epsilon_1$ is rather shallow, so this magnitude of the mini-gap is realised over
a significant interval of detunings $\delta \approx 0.05\ldots0.3\epsilon_F$. At larger detuning, the energy scale of the subgap states decreases linearly 
with $\delta$ and becomes exponentially small as $\delta\to2\epsilon_F$.
To conclude with an example, for a trapped potassium gas with $\epsilon_F=10$kHz we predict that the mini-gap can be of the order of $100$Hz at a detuning
frequency of $\delta\approx 1500$Hz.

\begin{figure}[tbhp]
\begin{center}
\includegraphics[width=0.8\columnwidth]{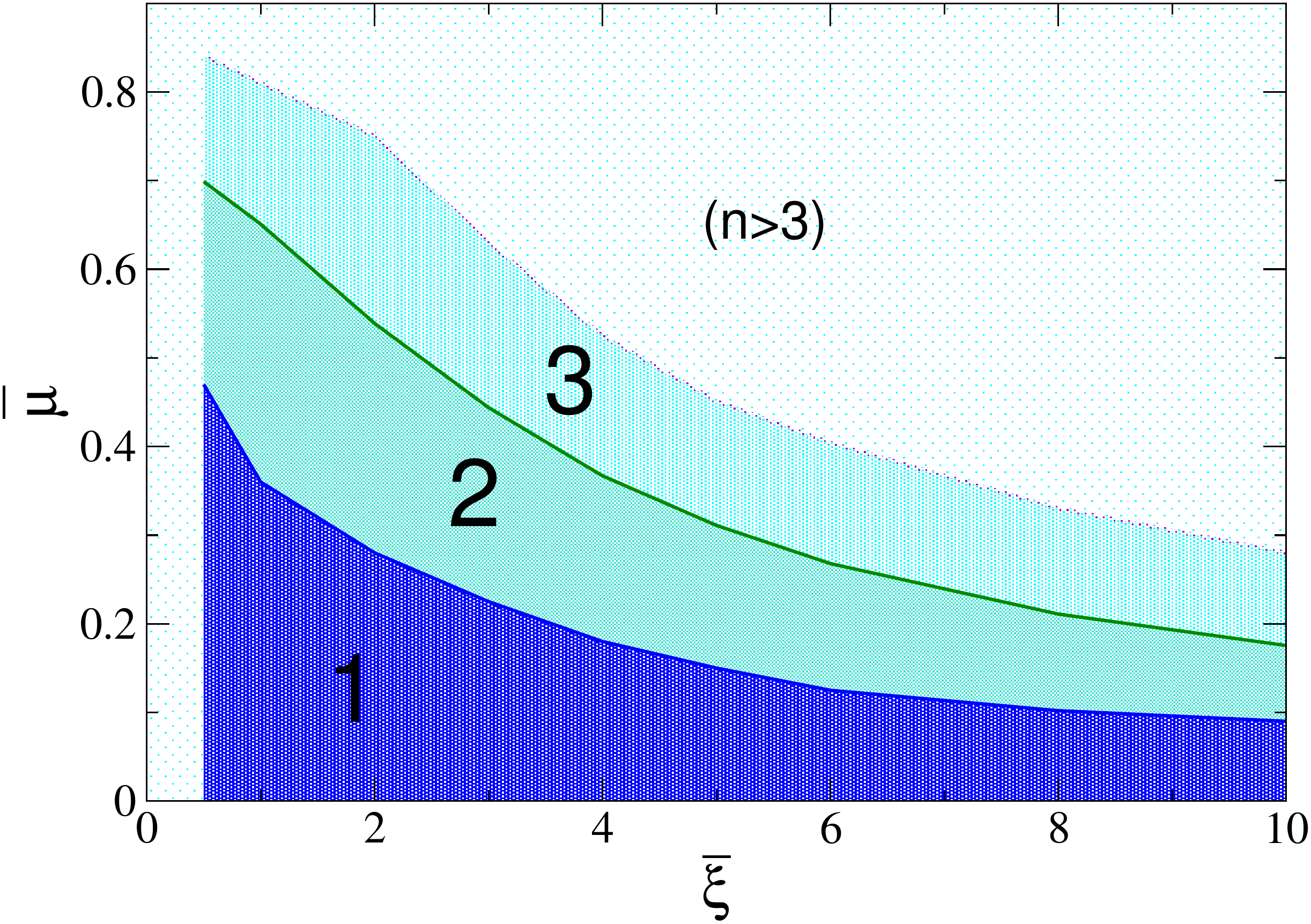}
\caption{(color online) Summary of how the number of Caroli--de Gennes--Matricon subgap states varies with the size of the vortex cores $\bar\xi$ and the chemical potential $\bar\mu$ [in dimensionless units according to Eq.~(\ref{eq:rescalings})]. Typical type II superconductors yield $\bar\xi\simeq1$, while $\bar\xi$ behaves according to Eq.~(\ref{eq:mudim}) in cold atomic gases.}
\label{fig:map}
\end{center}
\end{figure}

This energy scale for the mini-gap is still rather small, and we now consider mechanisms to further increase $\epsilon_1$. As energies are scaled with
$\mathcal{S}^{-1}$, we can read off from Eq.~(\ref{eq:scaleS}) how this can be achieved. In particular, decreasing the perpendicular confinement length of the
gas $\ell$ represents a simple means to enhance $\epsilon_1$ slightly. Further improvements can only be achieved by using a different Feshbach
resonance with larger coupling constant $g$ and smaller $c_2$. However, as very small $c_2$ will result in large vortex cores, this parameter should not
be smaller than about $1$. To visualise the scaling of energies as the vortex size $\bar\xi$ is varied, Fig.~\ref{fig:map} indicates the dependency the number 
of bound states below the bulk gap depends on the dimensionless parameters of the problem.
Besides increasing the overall energy scale, reducing $\mathcal{S}$ also shifts the maximum mini-gap towards larger values of the 
detuning, which would be easier to stabilise experimentally. 

In cold atomic gases, the presence of the subgap states can be probed experimentally by using RF spectroscopy, in the same way that had been 
proposed as a probe for the zero modes.\cite{Grosfeld07} The idea is to probe the amplitude for resonant absorption of an RF photon leading to a 
transition to a different hyperfine state, and therefore projecting the affected atom out of the spin-polarized Fermi gas. The bulk signal amounts 
to an absorption edge at $\hbar \omega = E_\text{hyperfine} - \mu + \Delta_B$, and the presence of the zero modes in vortex cores results amounts to the
addition of a series of linearly spaced peaks starting at the lower energy $\hbar \omega = E_\text{hyperfine} - \mu$ whose intensity decays on approaching
the absorption edge.\cite{Grosfeld07} The Majorana mode yields a sequence of peaks by coupling to different states of the continuum. For trapped gases,
these are spaced by the trap-frequency $\omega_\perp$. Finite energy subgap states result in additional series of absorption peaks occurring. In order for 
these different signals to be well separated, it is required that the spacing of peaks in each series be small compared to the typical energy gap between subgap states, i.e., that the trap frequency be smaller than the typical spacing of the subgap states.

\section{Conclusions}
\label{sec:conclusions}

To summarize the main results of this paper, we have shown that the presence of subgap states complicates the use of $p$-wave superfluids
as the medium underlying a topological quantum computer.  While topologically protected operations are still possible, as information remains
localized at a single vortex core even as transitions involving the Majorana zero mode and finite energy subgap states occur, the state of a qubit
depends additionally on the parity of the number of excitations of the subgap states. This disqualifies some known strategies for the read out of 
such quantum bits, especially for neutral superfluids which are the main focus of this paper.

As the number of subgap states scales roughly as the ratio of the bulk gap to the chemical potential, we argue that the regime of small chemical 
potential, or strong coupling should be most suitable to overcome these issues and maximize the mini-gap to the first subgap state. 

We note in passing that unconventional realizations of $p$-wave superconducting order where the magnitude of the order parameter is set 
externally\cite{FuKaneProximity2008,SauProximitySC2010, AliceaSandwich} may require different strategies to optimize the topological protection 
with regard to the subgap states.

In particular, we study the case of atomic Fermi gases, where this regime can be reached when tuning the system
close to a Feshbach resonance. We solve the Bogoliubov de Gennes equations for strong interactions in narrow Feshbach resonances, 
and calculate how the relevant system parameters, namely the chemical potential and the coherence length, depend on the detuning. 
Unlike typical type II superconductors in the BCS regime, the size of vortex cores can be tuned in cold atomic gases, allowing large vortices.

As Kopnin and Salomaa's perturbative solution of the vortex states is not valid in small chemical potential, we confirm numerically that the energy 
of the subgap states remains finite in small chemical potential; the subgap states are ultimately absorbed into
the continuum above the bulk gap at chemical potentials of $\mu\sim 0.5 \,m\Delta_0^2$. Among the most interesting aspects of the structure of the
vortex states we considered, we note a small splitting between the states associated with vortices and antivortices. In large vortex cores, we characterize
the nature of additional subgap states with a non-zero radial quantum number. Unlike the case of $s$-wave superconductors, these branches
are not symmetric under reversal of angular momentum.
In the particular case of $^{40}$K gases, we find that the optimal value of the detuning $\delta \approx 0.124\epsilon_F$ yields a maximal subgap energy
of the order of $0.01\epsilon_F$. Mechanisms for increasing this energy scale were discussed.

The authors would like to thank L.~Radzihovsky and S.~H.~Simon for discussions.
GM acknowledges support from Trinity Hall Cambridge, as well as from an ICAM fellowship and the ICAM Branches Cost Sharing Fund. 
We also acknowledge support from EPSRC grant no.~EP/F032773/1 (NRC) and by NSF grants DMR-0449521 and PHY-0904017 (VG).


\appendix

\section{$p$-wave Feshbach resonance parameters for $^{40}$K}
\label{sec:experimentalC2}
In this appendix, we derive the parameters of the $p$-wave Feshbach resonance in $^{40}$K. To do this, we need to restore all physical 
units, thus unlike in the rest of this paper where we set $\hbar=1$, here we restore $\hbar$ in every formula. 

$P$-wave superfluids with a Feshbach resonance are characterized by two parameters,\cite{GurarieRadzihovsky} the coupling constant $g$ and 
the ultraviolet cut-off momentum $\Lambda$. Instead of $\Lambda$,
it is more convenient to talk in terms of the dimensionless combination 
\begin{equation} 
c_2 =\frac{m^2}{3\pi^2} \frac{g^2 \Lambda}{\hbar^4}. 
\end{equation}
In the context of the current work, $c_2$  sets the coherence length of the superfluid, while $g$ sets the scale of the gap function $\Delta_0$. Here, we briefly review an argument from Ref.~\onlinecite{LevinsenPRA} describing
how $c_2$ and $g$ can be extracted from the detuning between closed channel bosons and open channel fermions in a two-channel model.
We then apply this method to the experimental data from Ref.~\onlinecite{TicknorBohn2004} to extract $c_2$ and $g$ for $^{40}$K.

The relevant physics is that of elastic scattering of two atoms, as described by the $p$-wave scattering amplitude $3 f_1(k) P_1(\cos\theta)$,
with the wave vector dependency
\begin{equation}
f_1(k)=\frac{k^2}{-\frac{1}{v}+ck^2 -ik^3}.
\end{equation}
Crucially, the scattering volume $v$ and the prefactor of the second order term $c$ can be extracted from experiments and from precise numerical modeling of the Feshbach resonance. On the other hand, we can relate $v$ and $c$ to the parameters $g$ and $c_2$. Indeed, we know that \cite{GurarieRadzihovsky}
\be \label{eq:parameters_raw} \frac{1}{v} = - \frac{6 \pi \hbar^2}{m g^2} \left( \epsilon_0-{\rm const} \right), \ c = - \frac{6 \pi \hbar^4}{m^2 g^2} \left(1+c_2 \right).
\ee
In turn, the energy $\epsilon_0$, which physically represents the Zeeman energy splitting between the open and closed channels can be quite generally written as
\begin{equation} \epsilon_0 = \alpha \mu_B B,
\end{equation}
where $\mu_B$ is the Bohr magneton and $\alpha$ is a dimensionless parameter controlling the Zeeman splitting and which we take to be close to $\alpha \approx 2$ (its precise value depends on the physics of Feshbach resonance).  This allows us to write
\begin{equation}  \label{eq:parameters}  \frac{d v^{-1}}{dB} = - \frac{6 \pi \alpha \mu_B \hbar^2 }{mg^2}, \ c = - \frac{6 \pi \hbar^4}{m^2 g^2}(1+c_2).
\end{equation}
Solving these for $g$ and $c_2$ gives
\be g^2 = - \frac{6 \pi \alpha \mu_B \hbar^2}{m \frac{d v^{-1}}{dB}}, \ c_2 = \frac{\alpha \mu_B m c}{\hbar^2 \frac{dv^{-1}}{dB}}-1.
\ee

In Ref.~\onlinecite{TicknorBohn2004}, $c$ and $v^{-1}$ are given as a function of $B$ for the $p$-wave Feshbach resonance occurring at  the magnetic field of $198.4$ Gauss for the hyperfine state $|f,m_f\rangle = | 9 / 2, -7 / 2 \rangle$ of potassium  $^{40}$K. Using their data, we can find the values of $c$ and $dv^{-1}/dB$ at 
the resonance and substitute them into (\ref{eq:parameters}). This gives
\be c_2 \approx 14.4, \ g \approx 1.40\cdot 10^{-46} {\rm J m}^{\frac 5 2}.
\ee

We can use the values found here to see at what detuning $\mu$ becomes of the order of $m \Delta_0^2/\hbar^2$, the units of energy controlled by $\Delta_0$. To do this, we rewrite (\ref{eq:mudim}) with $\hbar$ reintroduced [cf.~Eq.~(\ref{eq:mudim})]
\be  \frac{\hbar^2 \mu}{m\Delta_0^2} = 2 \pi (1+c_2) \frac{\hbar^4 \ell}{m^2 g^2} \frac{x}{1-x} = \mathcal{S} \frac{x}{1-x},
\ee
where we again introduced the parameter $\mathcal{S}$ from (\ref{eq:scaleS}), with $\hbar$ inserted as needed,
\be \mathcal{S} = 2 \pi (1+c_2) \frac{\hbar^4 \ell}{m^2 g^2}.
\ee
and as before $x=\delta/(2 \epsilon_F)$. 
Comparison with (\ref{eq:parameters_raw}) gives
\be \mathcal{S} = \frac{c \ell}{3}.
\ee
Substituting the value for the relevant parameter from Ref.~\onlinecite{TicknorBohn2004} (and using $\ell \sim 500$nm for the transverse width), we find
that $\mathcal{S} \approx 68.1$ and
\be  
\label{eq:Mu40K}
\frac{\hbar^2 \mu}{m\Delta_0^2} = 68.1 \cdot \frac x {1-x} 
\ee
We can estimate from here that in order to achieve the regime where $\mu$ is of the order of $m \Delta_0^2/\hbar^2$, i.e. roughly where the subgap states disappear, we need to keep \be x=\frac{\delta}{2\epsilon_F} \sim 0.014.
\ee
A more accurate calculation of the proposed target value of the detuning is given in section \ref{sec:exp_consequences} above, based on the evaluation of the full subgap spectrum.

\section{The subgap states for $\bar\mu\gg 1$: Kopnin and Salomaa's approach}
\label{sec:subgapAnalytic}

For completeness, this appendix includes a pedagogical summary of the solution to the BdG equations in the presence a single vortex (\ref{eq:BdGVortex}), as
first proposed by Kopnin and Salomaa.\cite{KopninSalomaa} We specialize to the case of vorticity $\kappa=1$, and point out that the BdG equations (\ref{eq:BdGVortex})
resulted from an Ansatz for a state with angular momentum $\langle L_z \rangle = m$ around the axis of the vortex. 
In particular, we highlight the consequences of the initial assumptions as we proceed through the 
analytical solution, with the assumptions being that 
\be
\bar\mu\gg 1,\quad  \bar E\sim 1.
\ee
The solution proceeds in three steps. First, we note that for large $\bar\mu$, the coupling of the equations (\ref{eq:BdGVortex}) for $u$ and $v$ is 
weak, and the solution is expected to be close to the decoupled equations given by the terms in square brackets only, which are solved by Bessel functions. Therefore, we proceed with the following Ansatz as the zeroth approximation
\be 
\label{eq:Ansatz_envelope}
u(r) = H_m^{(i)}(qr) \, f(r), \ v(r) = H_m^{(i)} (qr) \, g(r),
\ee
where we introduce $q= \sqrt{2\bar\mu} \gg 1$, and $H^{(i)}$ are the Hankel functions of type $i\in\{1,2\}$. To simplify notations, we explicitly state the following calculation for the case $i=1$.
For constant $f$ and $g$ we solve the parts of the equations in square brackets. More generally, we may allow these functions to be slowly varying, 
on distances of the order of $1$ (as opposed to the Hankel functions which vary over distances $\sim 1/q$). 

Substituting the Ansatz (\ref{eq:Ansatz_envelope}) into the equations (\ref{eq:BdGVortex}) and assuming that $r \gg 1/q$, the Hankel functions can be replaced by their 
asymptotic expansions $H(r)\sim e^{\pm i r}/\sqrt{r}$. In particular, their derivative is essentially proportional to the function itself times $i$, up to a small correction, 
namely $\derivX{r}H(qr)=iq H - \frac{1}{2r}H$. We find the following equations (no approximations, except the asymptotic expansion for the Hankel functions)
\begin{widetext}
\begin{eqnarray} \label{eq:B3}
- i q f' - \oh f''- \frac 1 {2r} f' + h^2 \left( i q g + g' + \frac{m}{r} g \right) + h h' g &=& \bar Ef,  \cr
- h^2 \left( i q f + f' - \frac{m}{r} f \right) - h h' f + i q g' + \oh g'' + \frac 1 {2r} g' &=& \bar E g.
\end{eqnarray}
\end{widetext}

This new set of equations (\ref{eq:B3}) can once more be analyzed in a perturbative approach. At the first level of approximation, we consider only the terms of order $q$ (in particular, 
dropping $E$ which is of the order of $1$)
\begin{eqnarray} \label{eq:fr}
- i q f' + h^2 i q g &=& 0 ,  \cr
- h^2 i q f + i q g'  &=& 0.
\end{eqnarray}
The solution to these equations yields a decaying behaviour at $r \rightarrow \infty$, which reads
\be
f = e^{- \int_0^r dr' h^2(r') }, \ g = -  e^{- \int_0^r dr' h^2(r') }.
\ee
Note that this approach does not capture the dependency of the localization length on the energy, as discussed above in section \ref{sec:asymptotics}.
To proceed with the perturbative solution of Eq.~(\ref{eq:fr}), we study corrections at the next order in $q$, taking the following 
Ansatz\cite{KopninSalomaa}$^,$\footnote{Ref.~\onlinecite{KopninSalomaa} has a typo in their version of Eq.~(\ref{eq:pert}). This Ansatz corresponds to the WKB approximation in the
third order, see Ref.~\onlinecite{LandauLifshitzVol3}.}
\be \label{eq:pert}
f \approx L(r)+ \frac{i \psi_1(r)}{q}, \ g \approx -L(r) + \frac{i \psi_2(r)}{q},
\ee
where we introduced a shorthand notation for the localization factor 
\be
L(r)=e^{- \int_0^r dr' h^2(r') }.
\ee
Substituting this last Ansatz into the equations \rf{eq:B3} and equating terms which are $q$-independent [the terms proportional to $q$ cancel
due to \rf{eq:fr}], one obtains
\begin{eqnarray}
\psi_1' - h^2 \psi_2 &=& \left( \bar E - \frac{h^4}{2} + h^2 \frac{m}{r} \right) L(r)  \cr
-\psi_2' + h^2 \psi_1 &=& \left( -\bar E - \frac{h^4}{2}-h^2 \frac{m}{r}  \right) L(r).
 \end{eqnarray}
 We look for a solution in the form
 \begin{eqnarray} \psi_1 &=& \alpha(r) \, L(r) + \beta(r) \, L^{-1}(r) \cr
  \psi_2 &=& - \alpha(r) \,  L(r) + \beta(r) \, L^{-1}(r)
  \end{eqnarray}
 The resulting equations for $\alpha$ and $\beta$ are a coupled system of differential equations
 \begin{eqnarray}
\alpha' L(r) + \beta' L^{-1}(r) &=& \left( \bar E - \frac{h^4}{2} + h^2 \frac{m}{r} \right) L(r) \\
\alpha' L(r) - \beta' L^{-1}(r) &=& \left( -\bar E - \frac{h^4}{2}-h^2 \frac{m}{r}  \right)  L(r), \nonumber
 \end{eqnarray}
which can be decoupled by taking the sum and difference to yield
 \begin{eqnarray}
 \alpha' &=& - h^4 \cr
 \beta' &=& \left( \bar E+ h^2(r) \,  \frac{m}{r} \right) L^2(r).
 \end{eqnarray}
 Integrating these equations, it follows that
 \begin{align}
 \alpha =& -\int_0^r dr' h^4(r') \\
 \beta =&- \int_r^\infty dr' \left( \bar E+ h^2(r') \,  \frac{m}{r'} \right) e^{- 2 \int_0^{r'} dr'' h^2(r'') } \nonumber.
 \end{align}
As the radial profile of the gap function $h(r)$ goes to $1$ at large distances this solution implies that $\psi_1$ and $\psi_2$ are well behaved at 
large distances (i.e., they vanish). 
 
Finally, we need to construct a solution to the full equations, which is regular at the origin.
This is achieved by matching boundary conditions on the set of solutions that we
have generated. The imaginary part of the Hankel functions which we used in the Ansatz 
(\ref{eq:Ansatz_envelope}) is divergent at the origin. A regular solution can be 
obtained by a superposition of the solutions expanded in $H^{(1)}$ and $H^{(2)}$. The solutions 
using $H^{(2)}$ are identical to the present discussion up to flipping the sign of $i$ 
in \rf{eq:B3}. For the superposition of the corresponding solutions to be regular at $r \rightarrow 0$,
the ensuing criterion is that $f(0)$ and $g(0)$ are the same for either of the two 
Hankel functions, such that the sum yields a Bessel function which is regular as $r \rightarrow 0$. 

Checking the Ansatz (\ref{eq:pert}), the condition is equivalent to $\psi_1(0) = \psi_2(0)=0$, as both functions enter with a factor of $i$ in \rf{eq:pert}, so they differ by a sign between 
the two Hankel functions. We always have $\alpha(0)=0$, but in addition it is required also that $\beta(0)=0$, which yields
 \be \label{eq:dr}
  \int_0^\infty dr \left( \bar E+ h^2(r) \,  \frac{m}{r} \right) e^{- 2 \int_0^{r} dr' h^2(r') }=0.
 \ee
 This yields the harmonic spectrum for the subgap states, $\bar E_m = - m \bar \omega_0$, with the spacing spelled out in Eq.~(\ref{eq:subgap_spacing}).
 One characteristic of this solution is that the energy of the subgap states depends crucially on the behaviour of the gap function near
 the vortex core. Indeed, if we had $h=$const, then $\bar \omega_0$ goes to infinity, as the integral over $r'$ in the numerator of \rf{eq:subgap_spacing}
 diverges in the origin. The only exception to this rule is the zero mode at angular momentum $m$=0. For it, as can be seen directly from \rf{eq:dr}, this 
 divergence is not relevant, indicating the exceptional topological protection of the zero-mode.

\section{Matrix elements between subgap states}
\label{sec:matrix_elements}
To make the nature of the matrix elements coupling the Majorana zero-modes and other subgap states more explicit, let us rewrite the single electron
creation operators $\hat a^\dagger$ in terms of the Bogoliubov transformed basis using the inverse of the unitary Bogoliubov transformation
\begin{align}
  \label{eq:expand_electron}
  \hat a^\dagger(\br) = & \sum_n \left[v_n(\br) \hat b_n + u_n^*(\br) \hat b_n^\dagger\right] \nonumber\\
  &+ \sum_j^{N_V/2}\left[  v_j^0(\br)\hat \Gamma_j + u_j^{0*}(\br) \hat \Gamma_j^\dagger \right]\\
  &+ \sum_{\nu}\sum_{m\neq 0}\left[ v_{\nu,m}(\br) \hat c_{\nu,m} + {u^*_{\nu,m}}(\br) \hat c_{\nu,m}^\dagger \right]\nonumber,
\end{align}
for an even number of vertices $N_V$. (In the case of odd $N_V$ an additional Majorana mode is found at the edge.) 
Here, we distinguish the contributions of the propagating Bogoliubov quasiparticles $\hat b_n$, the (pairs of) zero modes $\hat \Gamma_j$ 
and further subgap states $\hat c_{\nu,m}$ localized around the vortex $\nu$ and numbered also by their angular momentum $m$.
To show that both the left-hand and the right-hand side of Eq.~(\ref{eq:expand_electron}) are fermionic, we have expressed
the zero-modes in terms of fermionic operators obtained by pairing up individual localized Majorana zero-modes as 
\begin{equation}
\hat\Gamma_j=\frac{1}{2}[\hat\gamma_{2j}-i\hat\gamma_{2j-1}] ,
\end{equation}
with the associated Bogoliubov functions $u_j^0(\br)=u_{2j,0}(\br)+i u_{2j-1,0}(\br)$ and likewise for $v_j^0$. However, given
the property of the zero modes, $u_{j,0}^*=v_{j,0}$, their contribution can be rewritten as
\begin{align}
  \label{eq:expand_zero_modes}
   \sum_j^{N_V/2}\left[  v_j^0(\br)\hat \Gamma_j + u_j^{0*}(\br) \hat \Gamma_j^\dagger \right] = \sum_\nu^{N_V}v_{\nu,0} (\br) \hat\gamma_\nu,
\end{align}
which does not explicitly show the fermionic nature of the sum. We now apply the more compact notation of Eq.~(\ref{eq:expand_zero_modes}).
Using Eq.~(\ref{eq:expand_electron}) to express a scalar potential $\hat V = \int d^2r V(\br) \hat a^\dagger(\br)\hat a(\br)$ yields a 
bilinear form involving terms mixing $\hat b$'s, $\hat c$'s and $\hat\gamma$'s. The resulting expression includes, most 
prominently, the matrix elements between the states localized around the same vortices
\begin{align}
  \hat V \simeq & \sum_{\nu}\int d^2r \; V(\br) \\
  & \times \left[ v_{\nu,0}(\br) \hat\gamma_\nu + \sum_m \bigl(v_{\nu,m}(\br)  \hat c_{\nu,m} + u_{\nu,m}^*(\br)  \hat c_{\nu,m}^\dagger\bigr) \right] \nonumber\\
  & \times \left[v_{\nu,0}^*(\br)  \hat \gamma_\nu + \sum_m \bigl( u_{\nu,m}(\br)  \hat c_{\nu,m} + v_{\nu,m}^*(\br)  \hat c_{\nu,m}^\dagger\bigr) \right]  \nonumber.
\end{align}
As the zero-mode and subgap states all have support in a limited region around the vortex core, any
disorder potential that varies on the scale of the vortex is bound to induce sizeable matrix elements
for processes involving a single Majorana operator such as $\hat c^\dagger_\nu \hat\gamma_\nu$ or
$\hat\gamma_\nu \hat c_\nu$.
We have included explicitly only those processes which are coupling states at the same vortex;
other matrix elements are small assuming well separated vortices due to the exponentially decaying tails
of the vortex states (c.f. Eq.~\ref{eq:zero_mode}), as we discuss in more detail in the numerical
part of this paper.

\section{Form of the BdG equations on the sphere}
\label{sec:BdGSphere}
We briefly summarize the equations derived by Kraus {\it et al.}\cite{KrausPRL,KrausPRB} and transcribe them into
the dimensionless units used in this paper. When expressing the Bogoliubov-de Gennes equations for $p$-wave
superfluids, we choose a phase winding of $-2\pi$ for order parameter\footnote{The negative sign conforms with 
the choice in Ref.~\onlinecite{KrausPRB}, consistent with the pairing of the quantum Hall state at 
$\nu=5/2$ for $p_x-ip_y$ pairing.\cite{MollerSimon}} requires $q=- \frac{1}{2}$, corresponding to the 
basis functions in the presence of a single monopole flux;\cite{MollerSimon,KrausPRL,KrausPRB} the remaining quantum numbers $l,m$
span a complete basis of eigenfunctions of this problem with $l=|q|+s$, $|m|\leq l$ ($s=0,1,\ldots$). The BdG equations thus have the 
matrix form
\begin{equation}
  \label{eq:BdGSphere}
  \left(
    \begin{array}{cc}
      H_{(lm)(l'm')} &
      \Pi_{(lm)(\bar l'\bar m')} \\      
      \Pi_{(\bar l\bar m)(l'm')} &
      - H_{(\bar l\bar m)(\bar l'\bar m')}
    \end{array}
  \right)  
  \left(
    \begin{array}{c}
      u^n_{l'm'}\\
      v^n_{\bar l'\bar m'}
    \end{array}
  \right)
  =
  \bar E_n
  \left(
    \begin{array}{c}
      u^n_{lm}\\
      v^n_{\bar l\bar m}
    \end{array}
  \right),
\end{equation}
where the bar distinguishes indices that relate to the function $v$ and primed indices are summed over.
We study the vortex pair consisting of a vortex at the north pole and its antivortex at the antipode,\cite{KrausPRB}
with the vortex field 
\begin{equation}
\label{eq:vortex_field}
F_V(\theta,\phi) = \frac{\bar R/\bar\xi\sin\theta}{[1+(\bar R/\bar\xi\sin\theta)^2]^{\frac{1}{2}}} e^{i\phi},
\end{equation}
which conserves the angular momentum $m=\langle L_z\rangle$ as a good quantum number. The formulation
in Ref.~\onlinecite{KrausPRB} also introduces a pairing range $\xi_p$ which we take to be much smaller than all other length scales.
The off-diagonal (pairing) matrix elements of the resulting only couple the angular momenta $m=M$ and $\bar m = 1-M$, i.e., 
Eq.~(\ref{eq:BdGSphere}) separates into a block-diagonal form indexed by $M$ with remaining matrix indices $l$, $l'$.
The entries are,\cite{KrausPRB} with $D_l$ as in Ref.~\onlinecite{KrausPRB}
\begin{align}
  H_{(lm)(l'm')} = & \delta_{ll'}\delta_{mm'} \bar\mu\left(\frac{l(l+1)-1/4}{l_F(l_F+1)-1/4}-1\right)\\
  \Pi_{(lm)(\bar l'\bar m')} = & \delta_{m,1-\bar m'} \sqrt{2\bar\mu} \sqrt{\frac{1}{16\pi}(2l+1)(2\bar l+1)}\nonumber\\
  & \times [D_l+(-1)^{l+\bar l'}D_{\bar l'}]\sum_L\sqrt{2L+1}f_L^V\\
  & \times 
  \left(
    \begin{array}{ccc}
      l & \bar l' & L \\
      1/2 & -1/2 & 0
    \end{array}
  \right)
  \left(
    \begin{array}{ccc}
      l & \bar l' & L \\
      -m & m-1 & 1
    \end{array}
  \right),\nonumber
\end{align}
where the last line denotes the 3-J symbols for the coupling of angular momenta,
and $f_L^V$ are the expansion coefficients of the vortex field $F_V$ in regular spherical harmonics
\begin{equation}
  \label{eq:expansionfL}
f_L^V = \int d\Omega\, Y_{l,m=1}(\Omega)^* F_V(\Omega),
\end{equation}
Generally, the radius of the sphere needs to be chosen much larger than the size of the vortex core. 
For our numerical calculations, we used a maximum cut-off $l_\text{max} =400$. To minimize finite size effects, it is desirable to
choose the Fermi angular momentum $l_F$ as large as possible. Convergence of the numerical scheme depends
crucially on the ratio of $\bar R/\bar\xi$, and the maximal radius and Fermi angular momentum were determined in each case by the 
requirement of convergence of the eigenvalues.

For large $L$, it becomes difficult to calculate the expansion parameters (\ref{eq:expansionfL}) explicitly by numerical integration.
Instead, we fit the values of $f_L[\bar R,\bar\xi]$ for $L\gtrsim 100$ by a function of the form
\begin{equation}
  \label{eq:fit_expansion}
  f_L^{V,\text{fit}}[\bar R,\bar\xi]=a[\bar R,\bar\xi] e^{-b[\bar R,\bar\xi]L}/L^{c[\bar R,\bar\xi]},
\end{equation}
which yields an excellent approximation.

\bibliography{subgap}

\end{document}